\documentclass[sigconf]{acmart}
\AtBeginDocument{%
  }

\usepackage{graphicx}
\usepackage{float}
\usepackage{booktabs}
\usepackage{subcaption}
\usepackage{tabularx}
\usepackage{array}

\renewcommand\footnotetextcopyrightpermission[1]{}
\acmConference[Manuscript]{Manuscript}{2026}{}
\acmBooktitle{}
\acmDOI{}
\acmISBN{}

\begin{document}

\title{Cognitive Trajectory Modeling: Quantifying Human-AI Co-Creation through Cognitively Grounded Interaction Trajectories}

\author{Nicholas Davis}
\email{ndavis35@gatech.edu}
\affiliation{%
  \institution{Co-Creative AI Consulting}
  \city{Elyria}
  \state{Ohio}
  \country{USA}
}

\renewcommand{\shortauthors}{Davis}

\begin{abstract}
Co-creative AI research increasingly seeks methods capable of representing how interaction dynamics evolve through time. While many existing approaches focus on observable interaction characteristics, interaction metrics, behavioral coding schemes, or activity traces, these methods often struggle to capture higher-order interaction dynamics, including how collaborative processes reorganize, stabilize, regulate, and evolve through time. This paper introduces \textbf{Cognitive Trajectory Modeling (CTM)} as a cognitive theory of interaction dynamics that conceptualizes cognition, interaction, and creative processes as temporally organized trajectories unfolding across cognitively meaningful attractor landscapes. CTM builds upon the theoretical foundations of the Enactive Model of Creativity and Creative Sense-Making (CSM), revisiting the role of sense-making curves and cognitive trajectories in representing co-creative interaction dynamics. We formalize this perspective through the \textbf{Cognitive Trajectory Principle}, which states that temporal representations are only theoretically interpretable as cognitive trajectories when their underlying states possess directional cognitive meaning. Building on this principle, CTM generalizes the notion of cognitive trajectories beyond any particular coding scheme and provides a broader framework for modeling interaction dynamics through trajectories unfolding across meaningful attractor landscapes. We further distinguish cognitive trajectories from interaction traces and situate CTM within a broader hierarchy of cognitive, interaction, and domain dynamics. More broadly, we argue that understanding co-creative systems requires methods capable of modeling how cognition and interaction dynamics unfold through time. CTM provides a foundation for studying interaction dynamics across co-creative AI, human--AI collaboration, and the broader study of collective, distributed, and hybrid intelligence. More generally, CTM contributes to emerging efforts to quantify and model the temporal organization of human--AI interaction.

\end{abstract}

\ccsdesc[500]{Human-centered computing~Collaborative interaction}

\keywords{
Cognitive Trajectory Modeling,
Cognitive Trajectory Principle,
Interaction Dynamics,
Co-Creative AI,
Human-AI Collaboration,
Interaction-Centered Intelligence,
Quantified Co-Creation,
Computational Creativity,
Enactive AI,
Participatory Sense-Making,
Trajectory Analysis
}


\maketitle

\section{\textbf{Introduction}}

The rapid growth of co-creative AI has generated increasing interest in how collaborative interaction unfolds between humans and computational systems. Across computational creativity \cite{Kantosalo2016Modes,jordanous2016four},
human-computer interaction \cite{beaudouin2004designing},
design research \cite{dorst2011core},
creativity support tools \cite{Frich2019Landscape, Shneiderman2007CreativitySupport},
and interactive machine learning \cite{amershi2014, Fails2003},
researchers have increasingly shifted attention away from isolated system outputs and toward the dynamics of participation that emerge during collaboration itself \cite{Davis2015EnactiveModel, Frich2019Landscape, Kantosalo2016Modes}.

Rather than evaluating creativity solely through final artifacts, many researchers now seek to understand how creative outcomes emerge through ongoing interaction between agents, environments, tools, and evolving opportunities for action \cite{Karimi2018EvaluatingCreativity, guzdial2019interaction,lin2023ontology}. Recent work on creative activity traces similarly emphasizes the importance of documenting and interpreting the temporal structure of creative processes rather than treating creativity as a sequence of isolated state changes \cite{peng2026state}. These developments suggest a growing shift toward process-oriented accounts of creativity in which interaction itself becomes a primary object of analysis \cite{davis2026interaction}.

This shift has produced growing interest in methods capable of quantifying interaction dynamics through time. Researchers have explored a wide variety of approaches for analyzing co-creative interaction, including turn-taking behavior \cite{winston2017turn}, conversational structure \cite{bown2020speculative}, engagement metrics \cite{Karimi2018EvaluatingCreativity}, participation and interaction coding schemes \cite{Deshpande2023OCSM}, novelty and appropriateness assessments \cite{Deshpande2023OCSM}, temporal interaction analysis \cite{Davis2017CSM}, and, more recently, activity-trace and process-trace approaches that seek to document the structure of creative activity across time \cite{peng2026state,jo2026logs}. Collectively, these approaches reflect a broader shift away from evaluating creativity solely through final artifacts and toward understanding how creative outcomes emerge through ongoing interaction. These approaches reflect a broader recognition that creativity is not simply a property of isolated individuals or intelligent systems, but often emerges through processes of interaction, negotiation, exploration, adaptation, and collaborative sense-making \cite{Fischer2005SocialCreativity,Sawyer2007GroupGenius}.

Within this context, Creative Sense-Making (CSM) was introduced as a framework for quantifying interaction dynamics during open-ended co-creation \cite{Davis2017CSM}. Drawing from enaction \cite{stewart2010enaction,Varela1991EmbodiedMind}, participatory sense-making \cite{DeJaegher2007PSM}, distributed cognition \cite{Hutchins1995}, improvisation \cite{magerko2009empirical}, and embodied approaches to cognition \cite{DiPaolo2010Horizons,clark1997being}, CSM proposed that creative interaction could be understood as movement between distinct modes of cognitive participation \cite{Davis2015EnactiveModel}. Rather than treating interaction as a sequence of isolated observable events, the framework attempted to model how participants continuously regulate exploration, action, stabilization, and engagement with emerging possibilities across time \cite{Davis2017CSM}. This emphasis on evolving participation dynamics aligns with dynamical approaches to cognition that conceptualize cognitive activity as continuous trajectories through state spaces rather than discrete symbolic computations \cite{vanGelder1998}. The Sense-Making Curve was designed to reveal the temporal organization of cognitive dynamics by representing cumulative trajectories unfolding across a cognitively meaningful attractor landscape---that is, a structured space of cognitive possibilities containing regions of relative stability, transition, and reorganization.

More recent work in co-creative AI and interaction analysis has increasingly focused on observable interaction characteristics, including participation, novelty, and appropriateness \cite{Deshpande2023OCSM}, engagement \cite{Frich2019Landscape}, coordination and interaction structure \cite{Kantosalo2016Modes}, and other measurable dimensions of collaborative activity. These approaches have expanded the methodological toolkit available for studying co-creative systems and have contributed important insights into the observable dynamics of collaborative interaction.

At the same time, the growing diversity of interaction coding methods raises broader questions concerning how temporal representations of interaction should be interpreted. While many approaches generate visualizations, scores, or temporal summaries of interaction data, it remains unclear under what conditions such representations can be understood as trajectories of cognitive participation rather than descriptions of observable characteristics. More generally, the relationship between temporal representation, state-space structure, and trajectory interpretation remains under-theorized within co-creative AI research.

This paper argues that the interpretation of temporal interaction representations depends fundamentally upon the structure of the underlying state space from which they are generated. Building upon the Enactive Model of Creativity \cite{Davis2015EnactiveModel} and Creative Sense-Making \cite{Davis2017CSM}, we introduce the Cognitive Trajectory Principle, which proposes that temporal representations are only theoretically interpretable as cognitive trajectories when their underlying states possess directional cognitive meaning. Movement along a Sense-Making Curve from exploratory participation toward action-oriented cognition possesses directional cognitive meaning because each state corresponds to a distinct cognitive organization and the transitions between states represent interpretable changes in cognitive activity. The Cognitive Trajectory Principle therefore distinguishes temporal representations that merely summarize change through time from trajectories whose underlying states correspond to meaningful cognitive transformations.

This paper is guided by three broader research questions concerning the quantification of interaction dynamics in co-creative systems:

\begin{itemize}
    \item \textbf{RQ1. How can evolving interaction dynamics be represented as cognitive trajectories rather than isolated interaction events?} Traditional approaches often analyze interactions as discrete actions, behaviors, or observable characteristics. Cognitive Trajectory Modeling instead asks how participation evolves through time as a continuous process and how these evolving dynamics can be represented as trajectories within a cognitive attractor landscape.

    \item \textbf{RQ2. What conditions are necessary for cumulative interaction measures to generate theoretically interpretable cognitive trajectories?} While many interaction coding systems can be accumulated through time, not all accumulations produce meaningful trajectories. This question investigates the relationship between coding structure, directional state spaces, and trajectory generation, leading to the formulation of the Cognitive Trajectory Principle.

    \item \textbf{RQ3. How can cognitive trajectories be used to model, compare, and quantify participation dynamics in co-creative AI systems?} Beyond individual coding frameworks, this question explores how cognitive trajectories can serve as a general methodological foundation for studying interaction dynamics across human-AI collaboration, quantified co-creation, hybrid intelligence, and interaction-centered approaches to artificial intelligence.
\end{itemize}

\subsection{Contributions}

This paper makes five primary contributions to the study of co-creative AI, quantified co-creation, and interaction-centered approaches to intelligence.

\begin{enumerate}

\item \textbf{Cognitive Trajectory Modeling (CTM).}
We introduce Cognitive Trajectory Modeling (CTM) as a cognitive theory of interaction dynamics that conceptualizes cognition, interaction, and creative processes as temporally organized trajectories unfolding across cognitively meaningful attractor landscapes. CTM explains how cognitive trajectories give rise to interaction dynamics and how interaction dynamics contribute to the emergence of domain-level outcomes. Methodologically, CTM provides a family of trajectory-based approaches for representing, analyzing, and comparing the temporal organization of collaborative activity, shifting attention from isolated interaction events toward evolving patterns of stabilization, exploration, regulation, and reorganization through time.

\item \textbf{The Cognitive Trajectory Principle.}
We formalize the Cognitive Trajectory Principle, which states that temporal representations are only theoretically interpretable as cognitive trajectories when their underlying states possess directional cognitive meaning. This principle provides a foundation for distinguishing trajectory-generating state spaces from representations that merely summarize temporal variation.

\item \textbf{The Distinction Between Interaction Traces and Cognitive Trajectories.}
We develop a conceptual distinction between interaction traces and cognitive trajectories. Interaction traces describe observable characteristics, events, or metrics through time, whereas cognitive trajectories model how interaction dynamics evolve, stabilize, reorganize, and transform across time. This distinction clarifies an increasingly important methodological issue within co-creative AI and process-oriented approaches to creativity.

\item \textbf{Cognitive Dynamics, Interaction Dynamics, and Domain Dynamics.}
We introduce a three-level analytical framework distinguishing cognitive dynamics, interaction dynamics, and domain dynamics. Cognitive dynamics concern the organization of perception, attention, interpretation, and sense-making processes through time. Interaction dynamics concern the coordination, participation, regulation, adaptation, and co-regulation that emerge among interacting agents. Domain dynamics concern the evolution of artifacts, ideas, representations, and outcomes within a creative domain. This distinction provides a theoretical framework for understanding how cognitive trajectories contribute to interaction trajectories and how interaction trajectories contribute to the emergence of domain-level phenomena.

\item \textbf{Foundations for a Trajectory-Based Research Program.}
We position Cognitive Trajectory Modeling as a broader research program extending beyond Creative Sense-Making. We outline future directions involving multidimensional participation spaces, attractor landscapes, probabilistic trajectories, perceptual logic trajectories, and adaptive regulatory dynamics, providing a foundation for future work in co-creative AI, quantified co-creation, interaction-centered intelligence, and human--AI collaboration.

\end{enumerate}

\subsection{Paper Overview}

The paper proceeds in five stages. First, we revisit the theoretical foundations of the Enactive Model of Creativity and Creative Sense-Making, clarifying the role of directional participation states, mental models, sense-making curves, and cognitive trajectories in modeling creative interaction. Second, we introduce the Cognitive Trajectory Principle and examine how the structure of an underlying state space influences the interpretation of temporal interaction representations. Third, we develop a broader methodological distinction between directional and scalar state spaces, illustrating the conditions under which temporal representations can be interpreted as cognitive trajectories. Fourth, we trace the origins of Cognitive Trajectory Modeling (CTM) and present it as a general framework for representing participation as movement through cognitively meaningful attractor landscapes. Finally, we explore the implications of CTM for co-creative AI, quantified co-creation, human–AI collaboration, and interaction-centered intelligence, while outlining future directions involving multidimensional trajectories, attractor landscapes, probabilistic trajectories, perceptual logic trajectories, and adaptive regulatory dynamics.

\section{Historical Development of Creative Sense-Making and Trajectory-Based Analysis}

The development of Cognitive Trajectory Modeling (CTM) emerges from a broader research program investigating how cognition, creativity, and collaboration unfold through interaction over time. Rather than originating as a standalone framework, CTM represents the latest stage in a progression of theoretical, methodological, and computational developments spanning more than a decade.

Early work on \textbf{Perceptual Logic} (2011) introduced the idea that perception is not static but dynamically organized through evolving affordances and interaction patterns \cite{davis2011computing}. This work emphasized that opportunities for action emerge through ongoing engagement with an environment and can change over time as interaction unfolds \cite{davis2011computing}. Building on this perspective, \textbf{Human–Computer Co-Creativity} (2013) shifted attention from individual agents to interaction itself as the primary unit of analysis \cite{davis2013human}. Rather than treating computers as passive creativity support tools, this work explored how humans and computational systems could participate together in creative processes through collaborative interaction.

Related work on the \textbf{Enactive Characterization of Pretend Play} further developed this interaction-centered perspective by examining how meaning, imagination, and playful activity emerge through embodied participation and sense-making rather than through internal representations alone \cite{davis2015enactive}. Together, these projects contributed to an increasingly enactive understanding of cognition, creativity, and interaction.

This interaction-centered perspective was further formalized through the \textbf{Enactive Model of Creativity} (2014–2015), which framed creativity as an emergent process of participatory sense-making \cite{Davis2015EnactiveModel}. Creativity was understood not as the generation of ideas within isolated minds, but as a dynamic process arising through ongoing engagement among individuals, artifacts, environments, and computational collaborators. These foundations led to the development of \textbf{Creative Sense-Making (CSM)} (2017), which introduced a framework for analyzing how co-creative interaction unfolds through time \cite{Davis2017CSM}. CSM emphasized interaction dynamics, adaptation, and the temporal organization of creative engagement. As part of this work, \textbf{Sense-Making Curves} were introduced as a means of representing changes in interaction and creative participation across time \cite{davis2017quantifying,Davis2017CSM}. Subsequent work on \textbf{Quantified Co-Creation} (2018–2021) extended CSM into computational and empirical domains. Interaction patterns, collaboration dynamics, and creative trajectories were operationalized through quantitative models, enabling systematic analysis of co-creative processes across human–AI systems \cite{davis2025unlocking}.

The transition from descriptive trajectories to explanatory theory motivated the development of \textbf{Creative Trajectory Monitoring} (2021–2023). This approach began by generating creative activity traces, sense-making curves, and models of interaction dynamics. This research phase proposed that creative sense-making is organized as a directed and evolving trajectory characterized by coordination, negotiation, transitions between attractor landscapes, and periods of stability and change \cite{davis2025ai}.

\textbf{Cognitive Trajectory Modeling (CTM)} builds upon this lineage by providing a unified framework for representing, analyzing, and modeling cognitive trajectories in creative interaction. CTM integrates theoretical insights from enactive cognition and Creative Sense-Making with computational approaches to trajectory analysis, offering a general framework for understanding how cognition unfolds through time across human, social, and artificial systems.

\begin{figure*}[t]
  \centering
\includegraphics[
  width=\textwidth,
  height=.90\textheight,
  keepaspectratio]{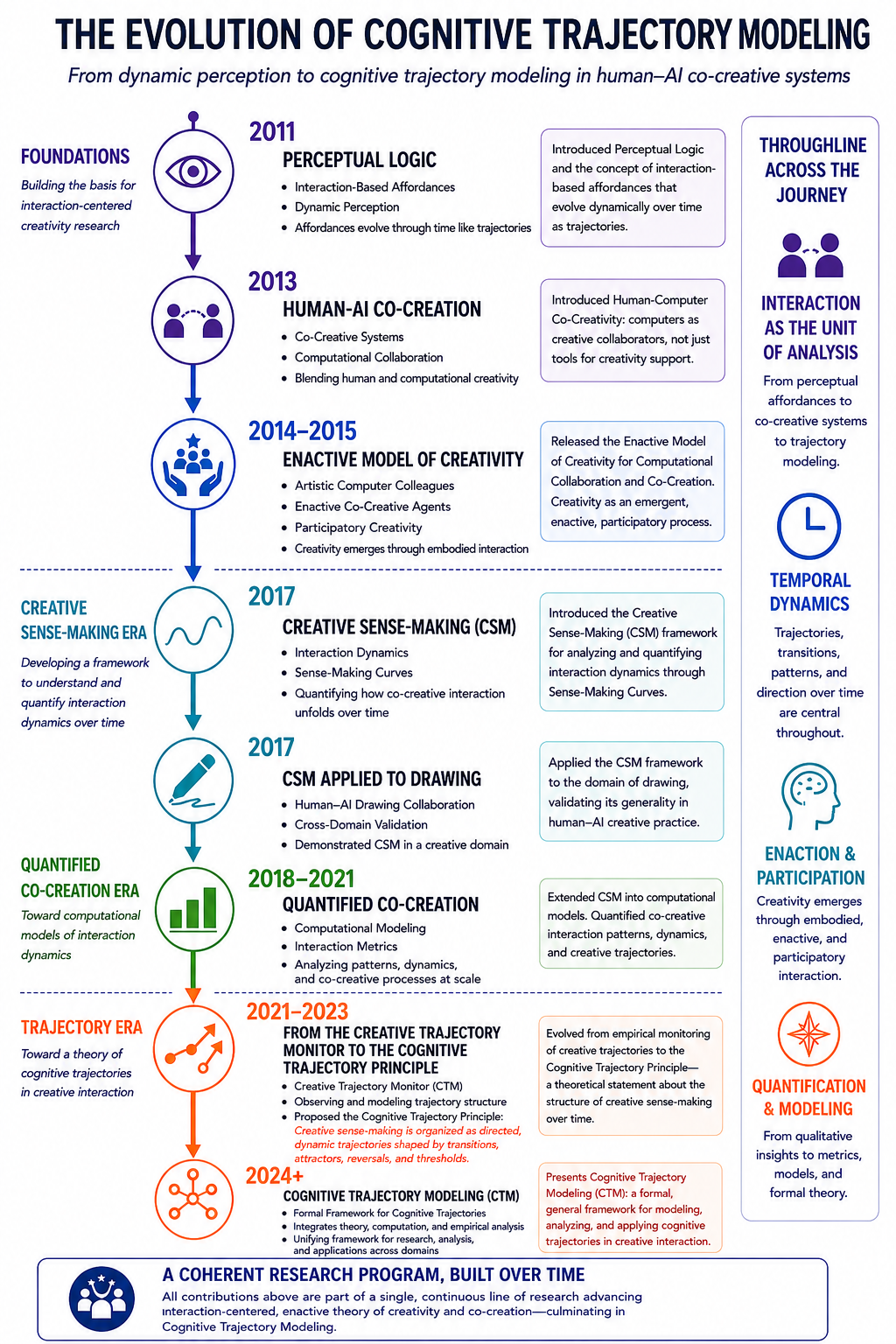}
  \caption{\textbf{Evolution of Creative Trajectory Modeling}}
  \label{fig:researchProgram}
  \Description{Timeline infographic illustrating the historical development of Cognitive Trajectory Modeling (CTM) from 2011 to the present. The figure presents a continuous research trajectory organized into three major eras: Foundations, Creative Sense-Making Era, Quantified Co-Creation Era, and Trajectory Era.

The Foundations begin with Perceptual Logic (2011), which introduced interaction-based affordances, dynamic perception, and the idea that affordances evolve through time as trajectories. Human-AI Co-Creation (2013) established interaction as the primary unit of analysis and framed computers as creative collaborators rather than passive tools. The Enactive Model of Creativity (2014–2015) introduced creativity as an emergent process arising through embodied interaction, participatory sense-making, and enactive co-creative agents.

The Creative Sense-Making Era begins with Creative Sense-Making (CSM) (2017), a framework for analyzing and quantifying interaction dynamics through Sense-Making Curves and evolving patterns of participation. During the same year, CSM was applied to human-AI drawing collaboration, providing cross-domain validation and demonstrating the framework in a creative co-creation context.

The Quantified Co-Creation Era (2018–2021) extended Creative Sense-Making into computational models, interaction metrics, and large-scale analysis of co-creative processes. This phase focused on computational modeling, interaction dynamics, and the quantitative study of creative interaction patterns.

The Trajectory Era (2021–2023) marks the transition from the Creative Trajectory Monitor to the Cognitive Trajectory Principle. Research shifted from empirical monitoring of creative trajectories to a theoretical framework proposing that creative sense-making is organized as directed, dynamic trajectories shaped by transitions, attractors, reversals, and thresholds.

The timeline culminates in Cognitive Trajectory Modeling (CTM) (2024–present), presented as a formal framework for representing, analyzing, and applying cognitive trajectories in human-AI co-creative systems. CTM integrates theoretical foundations, computational analysis, and interaction-centered approaches into a unified framework for understanding cognition, interaction dynamics, and creative processes as trajectories unfolding through time.

A sidebar identifies four themes that persist throughout the entire research program: (1) interaction as the primary unit of analysis, (2) temporal dynamics and trajectories as central explanatory concepts, (3) enaction and participatory sense-making as theoretical foundations, and (4) quantification and modeling as methodological goals. The figure argues that Cognitive Trajectory Modeling is the culmination of a coherent research program spanning more than a decade, integrating perceptual, cognitive, interactional, and computational approaches to creativity and co-creation.}
\end{figure*}

To situate CTM within its broader intellectual context, Figure \ref{fig:researchProgram} summarizes the development of the research program from Perceptual Logic (2011) through Human-AI Co-Creation, the Enactive Model of Creativity, Creative Sense-Making, Quantified Co-Creation, the Cognitive Trajectory Principle, and Cognitive Trajectory Modeling.

\section{\textbf{Creative Sense-Making as a Cognitive Framework}}

Creative Sense-Making (CSM) was originally introduced as a framework for quantifying interaction dynamics during open-ended co-creation \cite{Davis2017CSM,davis2017quantifying}. Unlike many observational coding approaches that focus primarily on describing visible behaviors, CSM emerged from a set of theoretical traditions concerned with cognition as an active \cite{noe2004action}, distributed \cite{Hutchins1995}, and participatory process \cite{DeJaegher2007PSM}. The framework was developed in response to a central challenge within creativity research and co-creative systems research: \textit{how can researchers study the evolving cognitive dynamics of creative interaction rather than merely its observable outcomes?}

The original Creative Sense-Making framework drew heavily from enactive cognition \cite{stewart2010enaction}, participatory sense-making, distributed cognition, and improvisation studies \cite{DiPaolo2010Horizons, DeJaegher2007PSM, Hutchins1995, Davis2015EnactiveModel}. Across these traditions, cognition is not understood as a process occurring solely within an isolated individual mind. Instead, cognition emerges through ongoing engagement between agents, environments, tools, constraints, and evolving opportunities for action \cite{Hutchins1995,Varela1991EmbodiedMind, gibson2014ecological}. Meaning is not merely produced internally and then expressed externally. Rather, meaning emerges through interaction itself \cite{DeJaegher2007PSM}.

This perspective is particularly important within co-creative contexts. During collaborative drawing, improvisation, design, movement, conversation, or human-AI interaction, participants continuously adapt to emerging conditions generated through the interaction process. Interaction dynamics are described as developing an autonomous organization that exerts influence on those involved in the interaction coupling \cite{DeJaegher2007PSM}. Creative action unfolds through cycles of exploration, response, interpretation, stabilization, and reorganization \cite{finke1992creative, Gabora2017}. As a result, understanding creativity requires more than describing observable actions. It requires understanding how participants cognitively engage with and contribute to the evolving interaction field---the emergent organization of meanings, affordances, constraints, and possibilities arising through ongoing interaction---whose organization both shapes and is shaped by ongoing participation \cite{Davis2015EnactiveModel, Sawyer2007GroupGenius}.

The Creative Sense-Making framework was developed precisely to address this problem. Rather than focusing exclusively on final creative artifacts or isolated interaction events, the framework sought to model how cognitive participation evolves through time. As stated in the original formulation, the goal was to develop a method for quantifying cognitive states, interaction types, dynamic trends through time, and sense-making patterns. This objective is significant because it places cognition—not merely behavior—at the center of the framework.

Within CSM, interaction codes do not function as descriptive labels attached to observable actions. Instead, they represent distinct modes of cognitive participation. The coding scheme was designed to capture how participants orient themselves toward the creative process at particular moments in time.

Three primary participation states were introduced:

\begin{itemize}
\item Unclamped participation (+1)
\item Clamped participation (-1)
\item Waiting or neutral participation (0)
\end{itemize}

These states were not intended as arbitrary numerical categories. Each state represents a different mode of cognitive engagement with the environment and the evolving creative task \cite{Davis2017CSM}. Unclamped participation refers to exploratory engagement. During unclamped states, participants actively inspect the environment, search for possibilities, generate alternatives, experiment with emerging ideas, and remain open to new affordances. The participant is not committed to a particular action trajectory but is instead engaged in exploratory sense-making. This mode reflects movement toward possibility generation, environmental inspection, and creative divergence. Clamped participation represents a contrasting mode of engagement. During clamped states, participants commit to specific actions, stabilize emerging structures, execute decisions, and narrow the space of possibilities into concrete forms. Rather than searching for alternatives, participants act upon selected directions. This mode reflects convergence, commitment, and stabilization within the creative process.

Waiting states occupy a transitional position between these two orientations. Waiting may involve observation, pause, reflection, inactivity, uncertainty, or temporary disengagement. Rather than indicating movement toward either exploration or stabilization, waiting functions as a neutral state within the coding framework.

Importantly, the numerical values associated with these states possess directional significance. The distinction between +1 and -1 is not equivalent to the distinction between low and high values on a rating scale. Instead, the values represent \textit{opposing cognitive orientations within the creative process}. Unclamped participation and clamped participation correspond to different directions of cognitive engagement \cite{Davis2017CSM}. This distinction is foundational to the framework. A participant moving into exploratory inspection of the environment is not simply exhibiting "more" or "less" of a characteristic. The participant is entering a different \textit{mode of sense-making}. Likewise, a participant committing to action is not merely increasing a behavioral score. The participant is shifting toward a different cognitive orientation.

Because these states possess directional meaning, they can be accumulated through time to generate interpretable cognitive trajectories. The coding scheme therefore functions as a representation of movement through a cognitive attractor landscape rather than a collection of isolated behavioral observations. This feature distinguishes Creative Sense-Making from many observational coding approaches. The framework was not designed to quantify the frequency of behaviors, assess degrees of creativity, or generate descriptive ratings of interaction quality. Instead, it was designed to model how participants move between exploratory and action-oriented modes of engagement as co-creative interaction unfolds \cite{Davis2017CSM}.

The significance of the coding scheme therefore lies not only in the classification of individual moments but in its ability to reveal larger patterns of cognitive organization across time. The directional structure of the coding system provides the foundation upon which the Sense-Making Curve becomes theoretically interpretable. Without this directional cognitive grounding, the cumulative trajectory would lose the interpretive structure that allows it to function as a representation of evolving participation dynamics.

Creative Sense-Making should therefore be understood not as an observational coding framework but as a cognitively grounded model of interaction dynamics. Its central aim is to represent the temporal organization of cognitive participation during co-creative interaction. The coding states are meaningful because they correspond to theoretically motivated modes of engagement, and the resulting trajectories are meaningful because they reveal how those modes evolve, accumulate, and reorganize through time.

\section{\textbf{The Theoretical Foundations of Creative Sense-Making}}

The directional participation states employed by Creative Sense-Making were not introduced as arbitrary coding categories. Rather, they were derived directly from the Enactive Model of Creativity \cite{Davis2015EnactiveModel}, which served as the theoretical foundation for the framework. Understanding why CSM uses directional states therefore requires understanding the underlying cognitive theory it was designed to operationalize.

\begin{figure}[t]
  \centering
  \includegraphics[width=\columnwidth]{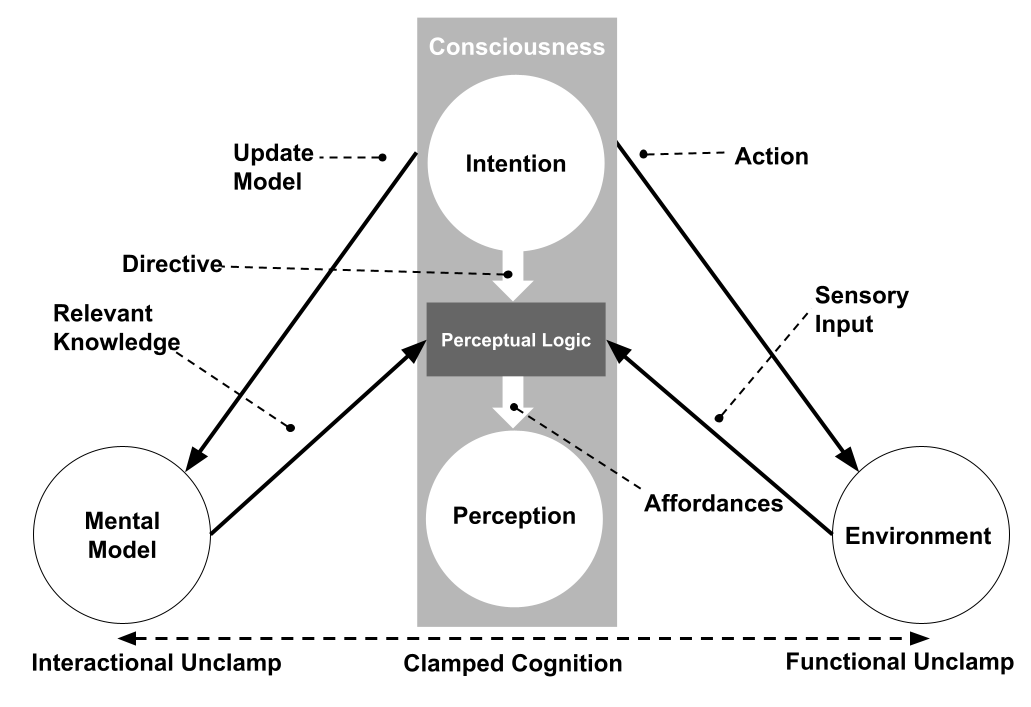}
  \caption{\textbf{The Enactive Model of Creativity (adapted from Davis et al., 2015).}
\normalfont{The model conceptualizes creativity as continuous regulation between unclamped exploration and clamped commitment and action. Creative Sense-Making operationalizes these directional modes as unclamped participation (+1), clamped participation (-1), and waiting (0), grounding the Sense-Making Curve in the underlying cognitive theory.}}

  \Description{
Diagram of the Enactive Model of Creativity (EMC). The model depicts creativity as a process emerging through ongoing interaction among a mental model, perception, intention, and the environment. At the center of the figure is a vertical cognitive process consisting of Perception at the bottom, Perceptual Logic in the middle, and Intention at the top, all contained within a region labeled Consciousness. Perceptual Logic functions as a regulatory mechanism that dynamically modulates affordance saliency based on current intentions and contextual factors.

On the left side is a Mental Model representing internally oriented cognition, including memory, imagination, reflection, simulation, and conceptual organization. On the right side is the Environment representing externally available affordances, sensory information, and opportunities for action. Bidirectional relationships connect the mental model, perception, intention, and environment, indicating continuous interaction between internal and external sources of meaning.

The model proposes that cognition can become increasingly unclamped toward either the mental model or the environment. Interactional unclamping occurs when attention shifts toward the mental model, increasing reflection, imagination, planning, and internal simulation. Functional unclamping occurs when attention shifts toward the environment, increasing affordance sensitivity, environmental inspection, exploration, and responsiveness to external information. Between these extremes lies clamped cognition, in which perception, intention, and action remain tightly coupled to ongoing interaction with the environment.

Creativity emerges through the continuous regulation of this relationship. Cognitive systems dynamically shift between internally oriented and externally oriented modes of participation, balancing reflection, imagination, exploration, perception, and action. The model therefore conceptualizes creativity not as isolated idea generation but as an emergent process of participatory sense-making arising through ongoing interaction among cognition, intention, perception, and the environment.
}
\end{figure}

The Enactive Model of Creativity emerged from enactive cognition, distributed cognition, and participatory approaches to creativity that view cognition as an active process of engagement with an environment rather than a purely internal process of idea generation. Within this perspective, creativity does not occur solely within the mind of an individual creator. Creativity emerges through continuous interaction between cognitive systems and the environments in which they participate. Creative cognition is therefore understood as a process of adaptive sense-making rather than the production of isolated ideas \cite{Davis2015EnactiveModel}.

A central claim of the Enactive Model of Creativity is that cognition continuously regulates its relationship to uncertainty, affordances, and opportunities for action. At any moment, a cognitive system may move toward greater exploration of the environment or toward greater commitment to particular actions \cite{davis2014building}. Creativity emerges through the ongoing modulation of these modes rather than through either mode alone \cite{Davis2015EnactiveModel}. The model therefore proposes a dynamic tension between exploration and action that organizes creative behavior across time.

Within the Enactive Model of Creativity, cognition may become increasingly unclamped. During these periods, attention shifts toward environmental inspection, possibility generation, experimentation, and exploration of alternative interpretations. The cognitive system becomes more sensitive to emerging affordances and novel opportunities for action. Unclamping therefore corresponds to exploratory engagement with the environment and the expansion of possible future trajectories.

Importantly, exploratory engagement may take two related forms. Cognition may move outward toward inspection of the external environment, searching for new affordances, materials, or opportunities for action. Alternatively, cognition may move inward toward revision and correction of existing mental models. In both cases, the cognitive system reduces commitment to existing action trajectories in order to increase sensitivity to new information and possibilities. The common feature of both processes is exploratory reorganization.

\begin{figure*}[t]
\centering

\begin{subfigure}{0.32\textwidth}
    \centering
    \includegraphics[width=\linewidth]{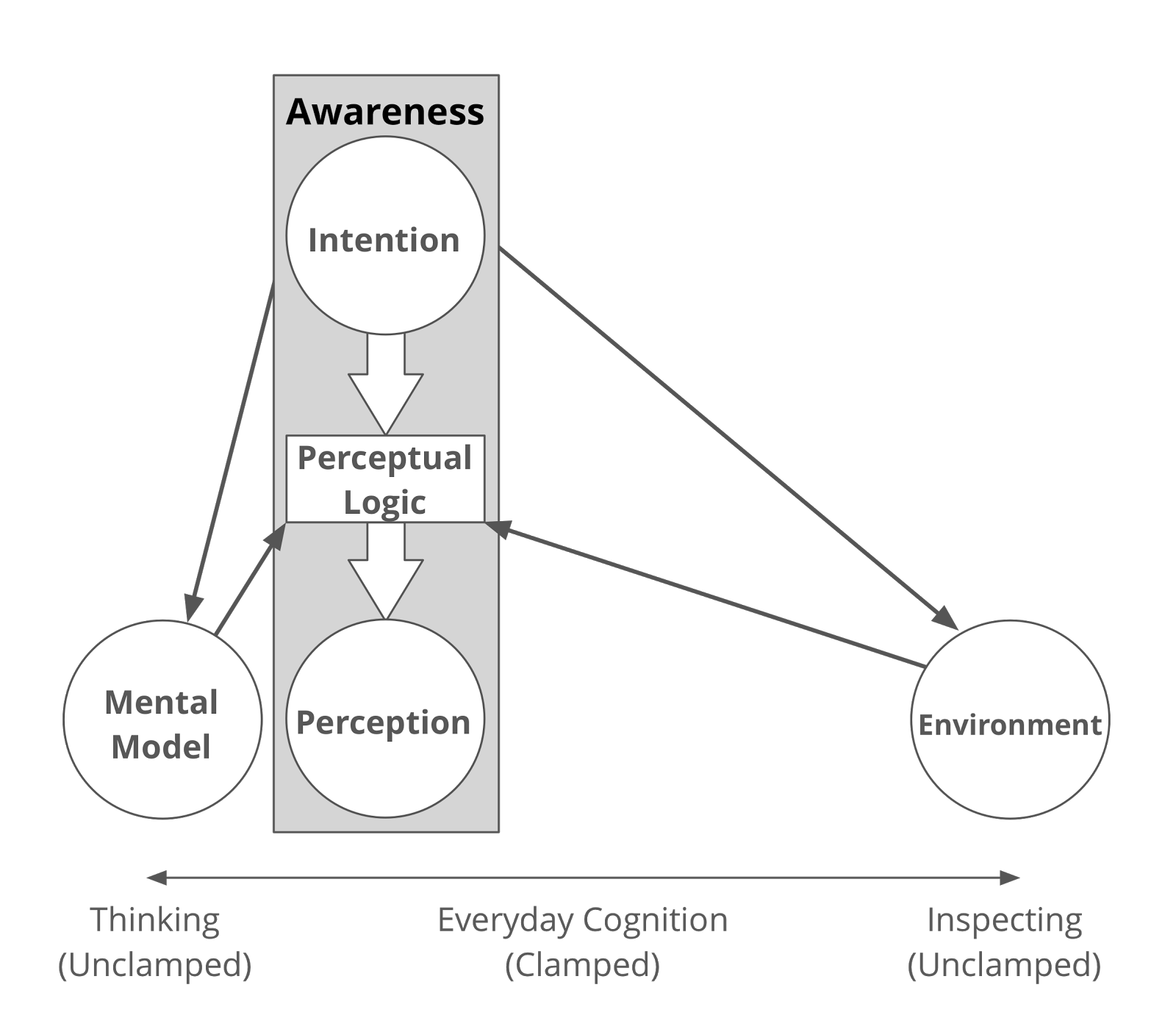}
    \caption{Model-Oriented Cognition}
\end{subfigure}
\hfill
\begin{subfigure}{0.32\textwidth}
    \centering
    \includegraphics[width=\linewidth]{images/EMC.png}
    \caption{Everyday Sense-Making}
\end{subfigure}
\hfill
\begin{subfigure}{0.32\textwidth}
    \centering
    \includegraphics[width=\linewidth]{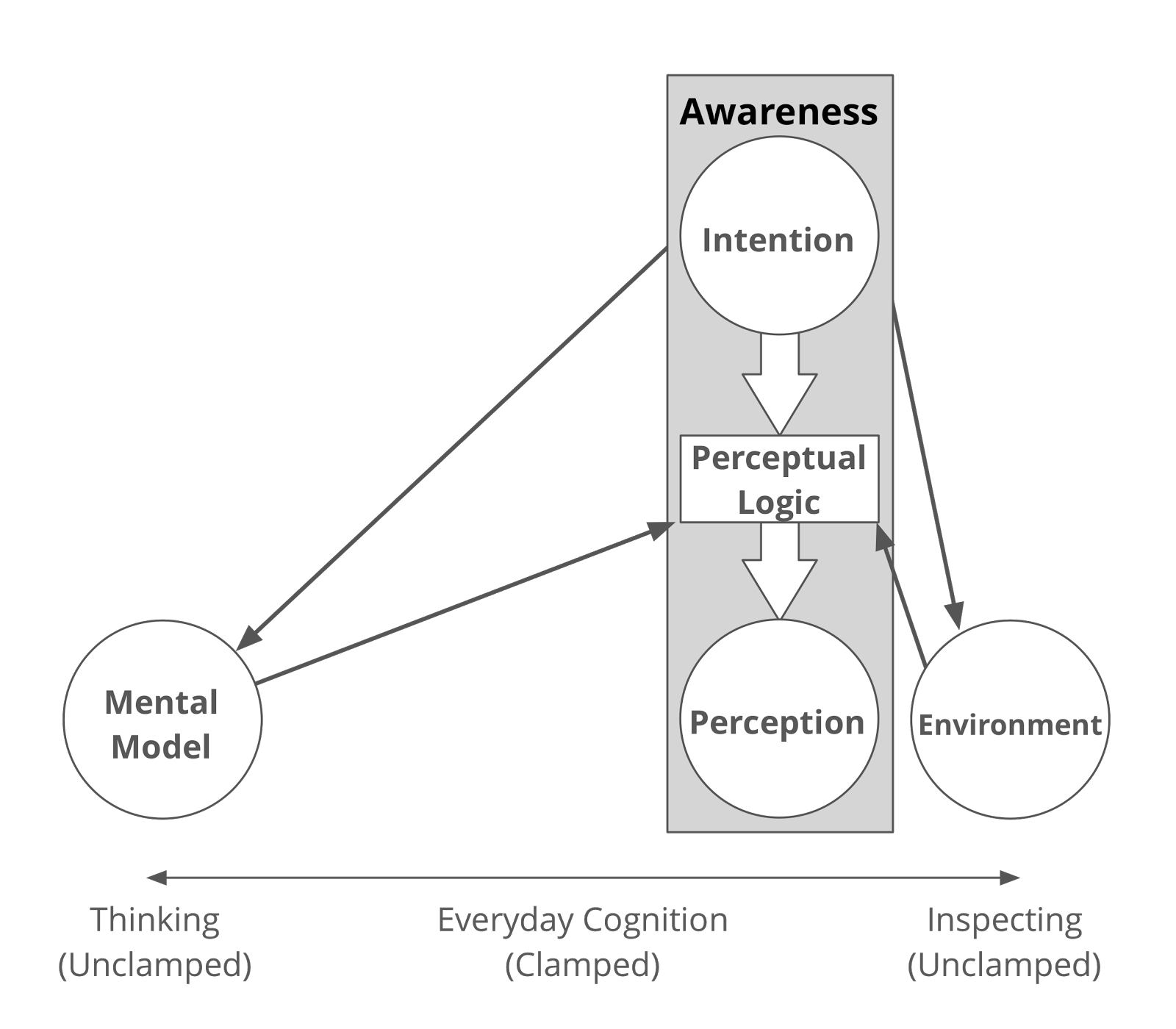}
    \caption{World-Oriented Cognition}
\end{subfigure}

\caption{\textbf{The Enactive Model of Creativity (EMC) as a process of cognitive modulation.}
\normalfont{Awareness shifts along a continuum between internally oriented and environmentally oriented cognition. Movement toward the mental model supports reflection, imagination, deliberation, and simulation, whereas movement toward the environment increases exploratory engagement, affordance sensitivity, and responsiveness to sensory information. Intermediate positions represent varying degrees of coupling between internal models and environmental input. Creativity emerges through the ongoing regulation of reflection, exploration, and action across this continuum.}}

\label{fig:cognitive_continuum}
\Description{
Three-panel diagram illustrating the Enactive Model of Creativity (EMC) as a continuum of cognitive regulation between internally oriented and environmentally oriented modes of participation.

Panel (a), Model-Oriented Cognition, depicts awareness shifted toward the mental model. Perception, intention, and perceptual logic remain present, but cognitive activity is primarily directed toward internal reflection, imagination, planning, simulation, memory, and conceptual organization. Environmental information exerts reduced influence on cognition. This state represents interactional unclamping, in which cognition becomes increasingly decoupled from immediate perception-action engagement with the environment.

Panel (b), Everyday Sense-Making, depicts an intermediate state in which awareness is balanced between the mental model and the environment. Perception, perceptual logic, and intention operate within an ongoing perception-action loop linking the cognitive system to environmental affordances. The mental model and environment both contribute information, and cognition remains coupled to ongoing interaction. This state represents clamped cognition and serves as the central mode of everyday sense-making.

Panel (c), World-Oriented Cognition, depicts awareness shifted toward the environment. Cognition becomes increasingly exploratory, affordance-sensitive, and responsive to sensory and interactional information. Environmental inspection and engagement dominate over internal reflection. This state represents functional unclamping, in which attention becomes increasingly directed toward opportunities for action and information available in the environment.

Taken together, the three panels illustrate creativity as a process of continuous cognitive modulation rather than a fixed mental state. Awareness moves along a continuum between internally oriented reflection and externally oriented exploration. Creativity emerges through the ongoing regulation of this relationship, enabling cognitive systems to dynamically balance imagination, planning, perception, affordance discovery, exploration, and action during creative activity.
}
\end{figure*}

Conversely, cognition may become increasingly clamped. During clamped periods, exploration is reduced and attention shifts toward execution, stabilization, implementation, and embodied action. Rather than generating additional possibilities, the cognitive system commits to selected affordances and transforms them into concrete behavior. Action becomes increasingly constrained by emerging goals, structures, and commitments. Changes in cognitive trajectories arise through shifts in how awareness is regulated relative to internal models and environmental information. Figure \ref{fig:cognitive_continuum} illustrates three characteristic organizations of this relationship. The Enactive Model of Creativity therefore describes creativity as a process of continuous regulation between exploratory and action-oriented modes of participation. 

Creative cognition is neither pure exploration nor pure execution. Instead, creativity emerges through ongoing movement between these complementary orientations. Novel possibilities are generated through exploratory engagement and subsequently stabilized through action. Stabilized structures may then be reopened to further exploration, producing cycles of divergence and convergence throughout the creative process.

The original Creative Sense-Making framework was designed as an operationalization of this theoretical model. The coding states were constructed specifically to capture these directional shifts in cognitive participation. Unclamped participation was assigned a value of +1 because it represented movement toward exploratory engagement, environmental inspection, and possibility generation. Clamped participation was assigned a value of -1 because it represented movement toward stabilization, commitment, and action execution. Waiting was assigned a value of 0 because it represented neither orientation. However, the waiting state plays an important structural role by preserving mutual exclusivity among cognitive state assignments and providing temporal continuity between active states, allowing the resulting representation to function as a \textit{continuous cognitive time series} rather than a collection of independent categorical observations.

The numerical values therefore reflected the structure of the underlying cognitive theory. They were not intended to function as scalar ratings or observational magnitudes. Instead, they represented directional positions within a cognitive participation state space derived from the Enactive Model of Creativity. This theoretical origin is important because it explains why the Sense-Making Curve was designed as a cumulative trajectory. The trajectory was intended to reveal how cognitive systems move between exploratory and action-oriented modes across time. The cumulative structure therefore follows directly from the Enactive Model of Creativity itself. Creative Sense-Making was not originally conceived as a behavioral coding framework that later adopted trajectories. It was conceived as a trajectory-based operationalization of an existing theory of creative cognition.

From this perspective, the directional participation states are not merely one possible implementation choice within CSM. They constitute the mechanism through which the Enactive Model of Creativity becomes measurable. The Sense-Making Curve derives its interpretability from this theoretical foundation. Without the directional structure inherited from the Enactive Model of Creativity, the framework would no longer represent the dynamics it was originally designed to capture. Directional states were inherited from the Enactive Model of Creativity because creativity itself was theorized as oscillation between exploratory and action-oriented modes of participation. Therefore removing directionality changes the ontology of the model.

\section{\textbf{Mental Models in the Enactive Model of Creativity}}

The inclusion of a mental model component within the Enactive Model of Creativity (EMC) may initially appear inconsistent with enactive approaches that reject internal representations as the primary basis of cognition \cite{Varela1991EmbodiedMind,thompson2010mind,clark1997being}.

Traditional cognitivist theories often assume that cognitive agents maintain persistent internal models of the world and rely on these representations to guide perception, reasoning, and action \cite{NewellSimon1976, Newell1980}. Influential approaches including the Physical Symbol System Hypothesis \cite{NewellSimon1976, Newell1980}, symbolic information-processing models \cite{Simon1979}, and classical cognitive architectures such as Soar \cite{LairdNewellRosenbloom1987} conceptualized cognition as the manipulation of internal symbolic structures that represent external states of affairs. Within this paradigm, intelligent behavior emerges from computational operations performed over these representations, allowing agents to reason, plan, solve problems, and act based upon internally maintained models of the world. In such accounts, cognition is frequently understood as the manipulation of internal representations that exist independently of ongoing interaction with the environment. In contrast, enactive and participatory approaches question whether cognition depends primarily upon pre-existing internal representations, emphasizing instead the ongoing dynamics of perception, action, and sense-making that emerge through interaction between agents and their environments \cite{Varela1991EmbodiedMind, DeJaegher2007PSM}.

The EMC employs the concept of a mental model in a fundamentally different manner. Rather than serving as a persistent representational structure, the mental model is understood as an emergent and temporary cognitive organization that arises during particular forms of cognitive activity. It is not assumed to continuously exist as a stable internal structure. Instead, it emerges when attention becomes increasingly oriented toward internally generated possibilities, simulations, memories, hypotheses, or conceptual relations.

Under normal conditions, cognition proceeds without requiring a mental model at all. Everyday cognition is primarily sustained through ongoing perception-action coupling between the cognitive agent and the environment. Action unfolds through direct engagement with affordances, environmental feedback, and continuous regulation of perception and behavior. In this state, cognition remains clamped within the interaction itself, and no explicit internal model is required to guide activity.

Mental models emerge when this coupling becomes partially relaxed or "unclamped" toward the mental domain. During moments of reflection, planning, imagination, creative ideation, explanation, or interactional breakdown, attention shifts away from immediate environmental engagement and toward internally generated structures. The resulting mental model functions as a temporary scaffold for sense-making. It provides a cognitive workspace through which the agent can explore possibilities, simulate actions, evaluate alternatives, and reorganize understanding.

Importantly, these structures are not treated as stored world models that persist independently through time. Rather, they are enacted in the moment as needed for ongoing sense-making. Once the relevant cognitive task has been completed and direct engagement with the environment resumes, the mental model recedes from attention and no longer serves as the dominant organizer of cognition.

This interpretation aligns with enactive accounts that view cognition as fundamentally grounded in embodied interaction while acknowledging that humans can temporarily direct attention toward internally generated structures. The mental model therefore functions less as a representation of the world and more as an emergent tool for thinking. It is a transient cognitive resource that arises during particular modes of engagement and dissolves when direct perception-action coupling is sufficient for ongoing activity. Stated succinctly, \textit{mental models are temporary, enacted cognitive scaffolds that emerge during breakdown, planning, imagination, or reflection and dissolve when perception-action coupling is restored.}

Within the EMC, mental models are therefore not treated as foundational cognitive structures. They are secondary phenomena that emerge under specific conditions of cognitive regulation. Everyday cognition remains grounded in perception-action coupling, while mental models appear primarily during reflective, imaginative, planning, or breakdown-related episodes in which cognition temporarily shifts toward internally oriented forms of sense-making. 

 Most cognition is perception-action coupling. Mental models only emerge temporarily when cognition becomes decoupled from immediate interaction and needs a scaffold for reflection, simulation, explanation, planning, or reorganization. When perception-action coupling becomes insufficient, cognition can temporarily construct schematic structures that function as tools for thinking. The model is not persistent. The model is enacted.

\section{\textbf{State Spaces, Attractor Landscapes, and Trajectories}}

\begin{figure*}[!t]
  \centering
  \includegraphics[width=\textwidth]{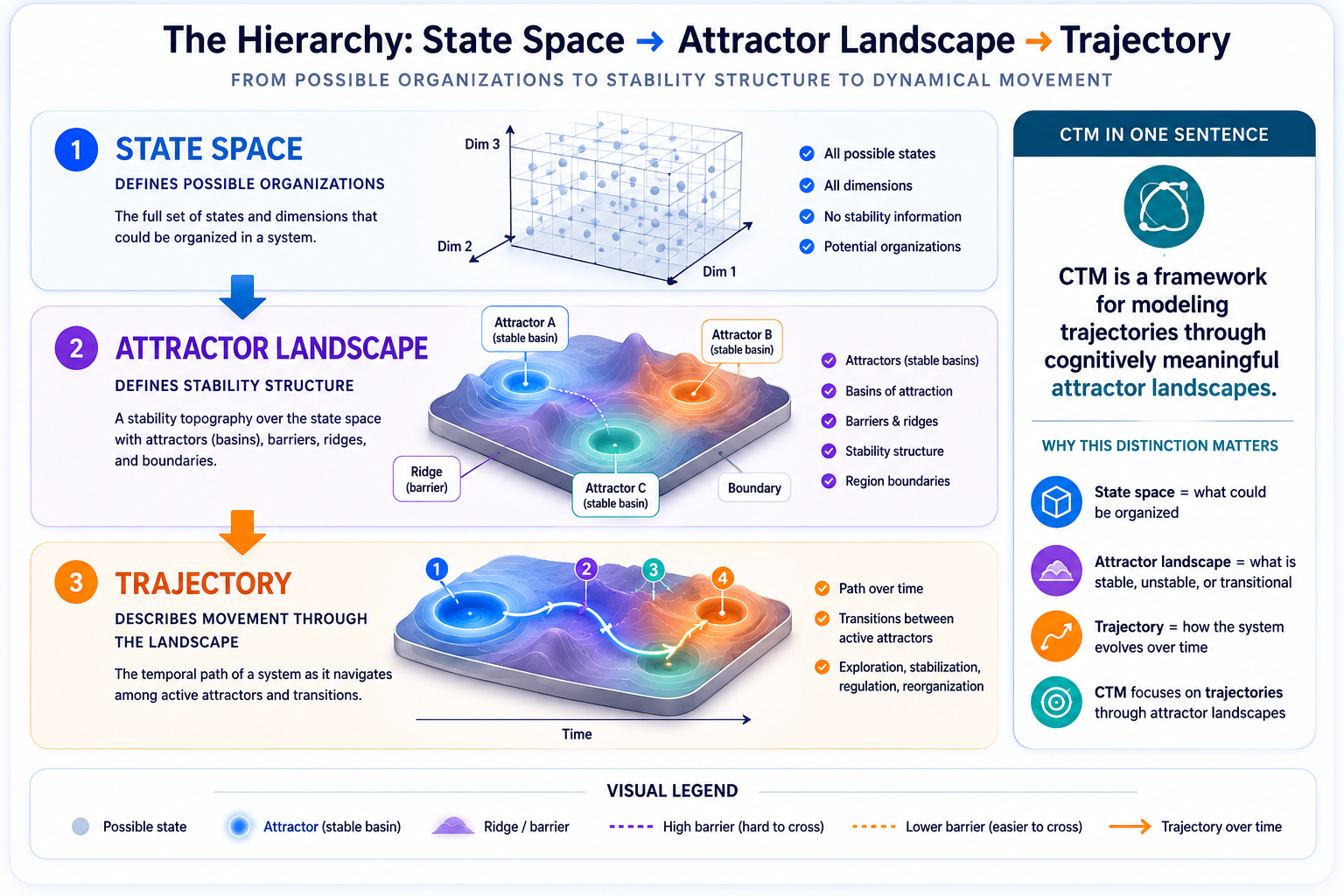}
  \caption{\textbf{The CTM hierarchy: from state spaces to attractor landscapes and trajectories.} 
  \normalfont{Cognitive Trajectory Modeling (CTM) conceptualizes cognition, interaction, and creative processes as trajectories unfolding within cognitively meaningful attractor landscapes. State spaces define the set of possible organizations available to a system by specifying its dimensions and potential states. Attractor landscapes impose stability structure on those possibilities by organizing regions into attractors (stable basins), barriers, ridges, and boundaries. Trajectories represent the temporal paths taken as a system explores, stabilizes, destabilizes, reorganizes, and transitions between attractors through time. From this perspective, CTM focuses not on isolated states or events, but on the evolving trajectories that emerge as cognition and interaction move across structured attractor landscapes.}}
\label{fig:stateSpace}
\Description{
Conceptual diagram illustrating the hierarchy underlying Cognitive Trajectory Modeling (CTM). The figure distinguishes three related levels of organization: state spaces, attractor landscapes, and trajectories.

At the top, a state space is represented as a multidimensional coordinate system containing all possible states and dimensions available to a system. The state space defines the set of possible organizations that may occur but contains no information about stability, transitions, or preferred regions.

In the middle, an attractor landscape imposes stability structure on the state space. The landscape contains three attractors labeled A, B, and C, represented as stable basins. Additional features include ridges, barriers, and boundaries that influence how movement occurs across the landscape. The attractor landscape determines which regions are stable, unstable, or transitional and shapes the organizational structure available to the system.

At the bottom, a trajectory is represented as a path moving through the attractor landscape over time. The trajectory begins in one attractor, transitions through intermediate regions, and eventually settles into other attractors. The path illustrates exploration, stabilization, destabilization, regulation, reorganization, and transitions between attractor basins.

A sidebar summarizes the core claim of Cognitive Trajectory Modeling: CTM is a framework for modeling cognition, interaction, and creative processes as trajectories unfolding through cognitively meaningful attractor landscapes. The figure emphasizes that state spaces define what could be organized, attractor landscapes define what is stable or unstable, and trajectories describe how systems evolve through time. Together these concepts form the foundation of Cognitive Trajectory Modeling and its representation of cognitive and interaction dynamics.
}
\end{figure*}

The concept of a cognitive trajectory originates from dynamical systems approaches to cognition, where cognitive activity is understood as evolving through structured organizational landscapes rather than progressing through isolated symbolic states \cite{beer2000dynamical, ThelenSmith1994}. Within this perspective, three related concepts—state space, attractor landscape, and trajectory—are useful for understanding Cognitive Trajectory Modeling, as shown in Figure \ref{fig:stateSpace}.

\begin{itemize}
    \item A \textbf{state space} defines the set of possible organizations available to a system. The dimensions of a state space specify the variables through which a system may vary and therefore determine the range of possible configurations that the system can occupy.
    \item An \textbf{attractor landscape} describes the stability structure imposed upon those possibilities. Certain regions of the landscape become more stable than others, forming attractors toward which the system tends to organize. Other regions may function as barriers, ridges, boundaries, or transitional zones that shape how change occurs through time.
    \item A \textbf{trajectory} is the actual path taken through the landscape as the system evolves. Trajectories reveal how a system explores possibilities, stabilizes within attractors, destabilizes existing organizations, and reorganizes into new patterns of activity.
\end{itemize}

From this perspective, cognitive trajectories are not simply sequences of observations. They are temporally organized pathways unfolding within structured cognitive attractor landscapes. This distinction provides the conceptual foundation for the Cognitive Trajectory Principle developed later in the article.

\section{\textbf{The Role of the Sense-Making Curve}}

The most distinctive feature of Creative Sense-Making is not the coding scheme itself, but the cumulative trajectory generated from that coding scheme. While the individual participation codes provide the foundational units of analysis, the primary theoretical object within the framework is the Sense-Making Curve \cite{Davis2017CSM}. The curve is not merely a visualization of coded interaction. It is the mechanism through which Creative Sense-Making models the temporal organization of cognitive participation during co-creative activity.

The significance of the Sense-Making Curve therefore extends far beyond its role as a visualization technique. The curve is the mechanism through which Creative Sense-Making transforms coded participation states into representations of evolving cognitive organization. Without the cumulative trajectory, the framework would be reduced to a collection of categorized interaction events. The larger patterns of exploration, stabilization, oscillation, transition, and regulation that define co-creative interaction dynamics would remain difficult to observe. \textbf{The cumulative trajectory is therefore not a supplementary analytic step. It is the theoretical core of the framework.}

Creative Sense-Making derives its explanatory power from its ability to generate interpretable cognitive trajectories that reveal evolving interaction dynamics, sense-making trends, and the temporal unfolding of cognition and perception through time \cite{Davis2017CSM}. The framework is ultimately concerned not with isolated behaviors, but with the evolving pathways through which cognition unfolds during interaction. The Sense-Making Curve provides a representation of those pathways, capturing patterns of stabilization, exploration, and transition as they emerge across the interaction process.

This distinction is critical because coding frameworks are often interpreted primarily as systems for categorizing events. From this perspective, the coding process is treated as the central analytical operation, while visualizations are viewed as secondary representations of already completed analysis. Creative Sense-Making operates differently. Within the framework, the individual codes possess meaning primarily because they contribute to the construction of a cumulative cognitive trajectory across time. The theoretical focus therefore shifts away from isolated interaction events and toward evolving participation dynamics. Rather than asking what cognitive state occurred at a particular moment, Creative Sense-Making asks how cognitive participation changes, accumulates, stabilizes, oscillates, and reorganizes across the unfolding interaction process. The Sense-Making Curve was developed specifically to reveal these larger temporal structures \cite{Davis2017CSM}.

\subsection{\textbf{Temporal Accumulation}}

The cumulative sum procedure transforms individual participation states into an evolving trajectory of interaction. At each moment, the current coded value is added to the accumulated trajectory generated by all previous states. When exploratory participation dominates over time, the trajectory rises. When action-oriented stabilization dominates over time, the trajectory falls. Periods of waiting produce temporary stabilization of the curve.

The resulting trajectory represents the evolving balance between exploratory and clamped modes of cognitive participation. This accumulation process is central because cognition unfolds through time rather than through isolated moments. A single exploratory action reveals relatively little about the larger organization of the interaction. Similarly, a single moment of commitment does not necessarily indicate broader stabilization dynamics. The cumulative trajectory reveals how participation states interact across extended temporal intervals. As a result, the Sense-Making Curve provides access to structures that are not visible at the level of individual codes alone. For example, the approach has the potential to quantify the following cognitive processes: 

\begin{itemize}
\item Exploratory phases
\item Exploitative phases
\item Oscillatory cycles
\item Transitions between modes
\item Stabilization periods
\item Collaborative regulation patterns
\item Interaction rhythms
\item Large-scale organizational dynamics
\end{itemize}

These structures emerge through temporal accumulation rather than through isolated observation. The cumulative curve therefore functions as a representation of evolving cognitive organization rather than a simple summary of coded events.

\subsection{\textbf{Cognitive Trajectory Generation}}

Within Creative Sense-Making, the curve is not intended to represent the amount of creativity, engagement, participation, or novelty present within an interaction \cite{Deshpande2023OCSM}. Instead, it represents how participants move through different modes of cognitive engagement as the interaction unfolds \cite{Davis2017CSM}. The resulting trajectory can therefore be interpreted as a path through an evolving landscape of cognitive participation. More broadly, this interpretation is consistent with dynamical systems approaches that analyze cognition and behavior as evolving trajectories through state spaces characterized by patterns of stability, transition, and self-organization \cite{Kelso1995}.

Rising segments indicate increasing exploratory engagement. Falling segments indicate increasing stabilization and action commitment. Oscillatory regions indicate dynamic movement between exploratory and exploitative participation. Plateaus indicate periods of relative stability or inactivity. The shape of the curve becomes analytically meaningful because it reflects the temporal organization of cognitive participation rather than merely the frequency of observable behaviors. The trajectory itself becomes the object of analysis. This shift is important because many interaction coding frameworks focus primarily on classifying events. Creative Sense-Making instead focuses on modeling how cognitive engagement evolves through time. The framework therefore moves beyond behavioral description toward trajectory-based representations of participation dynamics.

\subsection{\textbf{Continuous Cognitive Participation}}

An additional distinction concerns the temporal completeness of the coding system. Creative Sense-Making employs a set of mutually exclusive and collectively exhaustive participation states consisting of unclamped participation (+1), waiting (0), and clamped participation (-1). Because a participant must occupy one of these states at any given moment, the coding system produces a continuous temporal record of cognitive participation. The resulting trajectory reflects ongoing changes in cognitive organization across the entire interaction.

Many observational coding approaches operate differently. Rather than assigning participants to continuously occupied cognitive states, they assess characteristics of interaction such as novelty, appropriateness, engagement, or participation \cite{Deshpande2023OCSM}. These measures may be applied at high temporal frequencies and can generate dense temporal traces. However, they do not necessarily represent continuous state occupancy. Instead, they describe changing properties of interaction through time.

From the perspective of Cognitive Trajectory Modeling, this distinction is important because trajectories emerge not merely from temporal sampling frequency but from the continuous representation of movement between meaningful cognitive orientations. The existence of a waiting state further contributes to this continuity by allowing periods of observation, hesitation, reflection, and non-action to be represented within the trajectory itself. As a result, Creative Sense-Making generates a continuous record of cognitive participation rather than a sequence of independent interaction assessments.

\subsection{\textbf{The Cognitive Trajectory Principle}}

The theoretical logic underlying the Sense-Making Curve suggests a broader methodological principle that extends beyond Creative Sense-Making itself. We propose what may be called the Cognitive Trajectory Principle:

\begin{quote}
\textbf{The Cognitive Trajectory Principle states that a temporal representation of interaction dynamics is only theoretically interpretable as a cognitive trajectory when its underlying states possess directional cognitive meaning.}
\end{quote}

This principle identifies a necessary condition for generating meaningful cognitive trajectories through cumulative analysis. For a cumulative curve to represent movement within a cognitive attractor landscape, the coding values must correspond to theoretically meaningful orientations of participation. The values must indicate directional movement between distinct cognitive modes rather than merely differences in magnitude, frequency, intensity, or subjective assessment.

When directional cognitive meaning is present, temporal accumulation can generate interpretable trajectories that reveal the organization of participation across time. When directional cognitive meaning is absent, accumulation may still generate visual patterns, but the resulting curves no longer necessarily represent movement between attractor landscapes. This distinction is critical because cumulative visualizations can appear structurally similar while representing fundamentally different phenomena. In Creative Sense-Making, directional cognitive meaning is further reinforced through continuous state occupancy, ensuring that cognitive participation is represented across the entire interaction rather than only during observed events.

\begin{figure*}[!t]
  \centering
  \includegraphics[height=0.9\textheight]{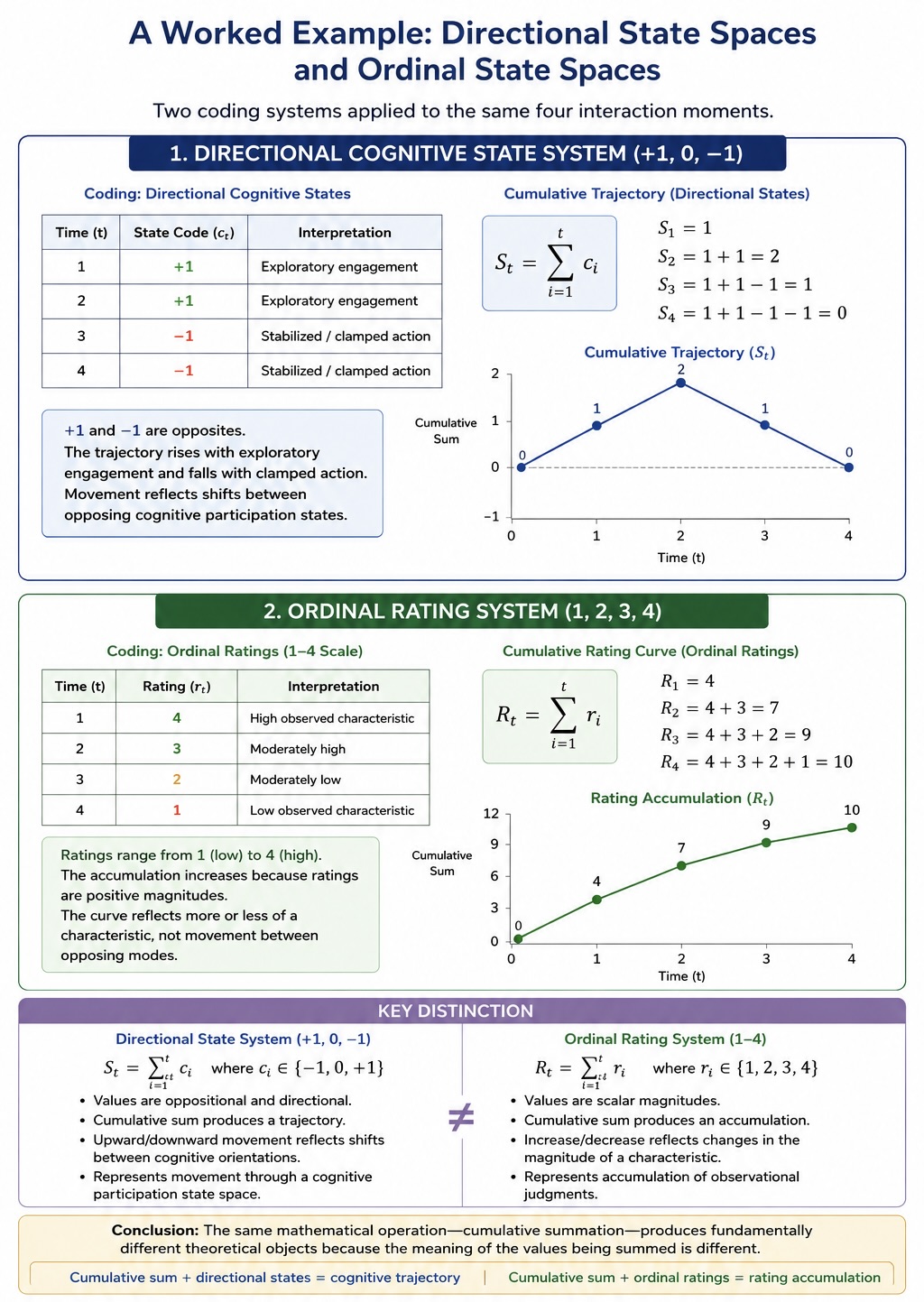}
  \caption{\textbf{ Illustrating the Cognitive Trajectory Principle.} 
  \normalfont{Two coding systems are applied to the same interaction sequence. The directional state system supports interpretation as movement through a cognitive attractor landscape, whereas the ordinal state system supports interpretation as variation in the magnitude of an observed characteristic. The figure is intended as a theoretical illustration of state-space structure rather than a representation of any specific interaction framework.
}}
\Description{
Worked example comparing directional cognitive state spaces and ordinal rating state spaces using the same sequence of four interaction moments. The figure demonstrates that identical mathematical operations can produce fundamentally different theoretical interpretations depending on the meaning of the underlying values.

The upper panel presents a directional cognitive state system using values of +1, +1, -1, and -1. The values represent opposing cognitive participation states, where +1 corresponds to exploratory engagement and -1 corresponds to stabilized or clamped action. A cumulative sum is computed across time, producing trajectory values of 1, 2, 1, and 0. The resulting graph rises during exploratory participation and falls during stabilized participation, illustrating movement through a directional cognitive state space. Because the values represent oppositional orientations, upward and downward movement reflects meaningful transitions between cognitive modes. The cumulative curve therefore constitutes a cognitive trajectory.

The lower panel presents an ordinal rating system using values of 4, 3, 2, and 1. These values represent observational ratings ranging from high to low levels of a characteristic. Applying the same cumulative summation procedure produces values of 4, 7, 9, and 10. The resulting graph increases monotonically because all ratings are positive magnitudes. The curve reflects the accumulation of ratings rather than movement between opposing states. Changes in the slope indicate differences in the magnitude of observed characteristics rather than transitions between cognitive orientations.

A comparison section highlights the distinction between the two systems. In the directional state system, values are oppositional and directional, cumulative summation produces a trajectory, and movement through the resulting curve reflects shifts between cognitive participation states. In the ordinal rating system, values are scalar magnitudes, cumulative summation produces an accumulation curve, and changes reflect variations in the strength of an observed characteristic rather than movement through a cognitive state space.

The central conclusion of the figure is that the same mathematical operation, cumulative summation, produces fundamentally different theoretical objects depending on the meaning of the values being summed. Cumulative summation applied to directional cognitive states yields a cognitive trajectory, whereas cumulative summation applied to ordinal ratings yields rating accumulation. The figure illustrates the Cognitive Trajectory Principle that trajectories require directional cognitive meaning in the underlying state space rather than merely temporal ordering or numerical magnitude.
}
  \label{fig:regCenteredAI}
\end{figure*}

\section{\textbf{Cognitive Trajectory Modeling: A Framework for Interaction Dynamics}}

Cognitive Trajectory Modeling identifies a family of methods concerned with representing interaction as movement through temporally organized attractor landscapes. These approaches differ from interaction-analysis methods that focus primarily on observable characteristics, behavioral coding, or descriptive interaction measures \cite{Deshpande2023OCSM}. This distinction suggests the need for a broader conceptual category capable of identifying and organizing methods that attempt to represent cognition through evolving interaction trajectories rather than through isolated observations alone. To address this need, we introduce the concept of \textbf{Cognitive Trajectory Modeling (CTM). }Cognitive Trajectory Modeling refers to a family of methodological approaches that attempt to model the temporal organization of cognitive participation through evolving trajectories generated from directional state representations. Rather than focusing primarily on static observations, aggregate measurements, or isolated interaction events, CTM approaches treat cognition as a dynamic process unfolding through time. The primary object of analysis is not the individual observation. The primary object of analysis is the trajectory.

\subsection{\textbf{From State Classification to Trajectory Modeling}}

Many interaction analysis frameworks focus on classification. Researchers identify categories, labels, ratings, behaviors, events, or interaction characteristics and then analyze their frequency, distribution, or occurrence patterns. Within such approaches, individual observations often function as the central unit of analysis.

Cognitive Trajectory Modeling operates differently. While CTM frameworks may also employ coding procedures, the purpose of coding is not merely classification. Instead, coding functions as a mechanism for constructing interpretable trajectories that reveal larger patterns of cognitive organization across time. The significance of a coded event emerges not only from its local classification but from its role within an evolving participation pathway. From this perspective, cognition is not modeled as a collection of isolated states. It is modeled as evolving participation across a structured cognitive attractor landscape. The trajectory becomes the primary analytical object.

\begin{figure*}[!t]
  \centering
  \includegraphics[width=.8\textwidth]{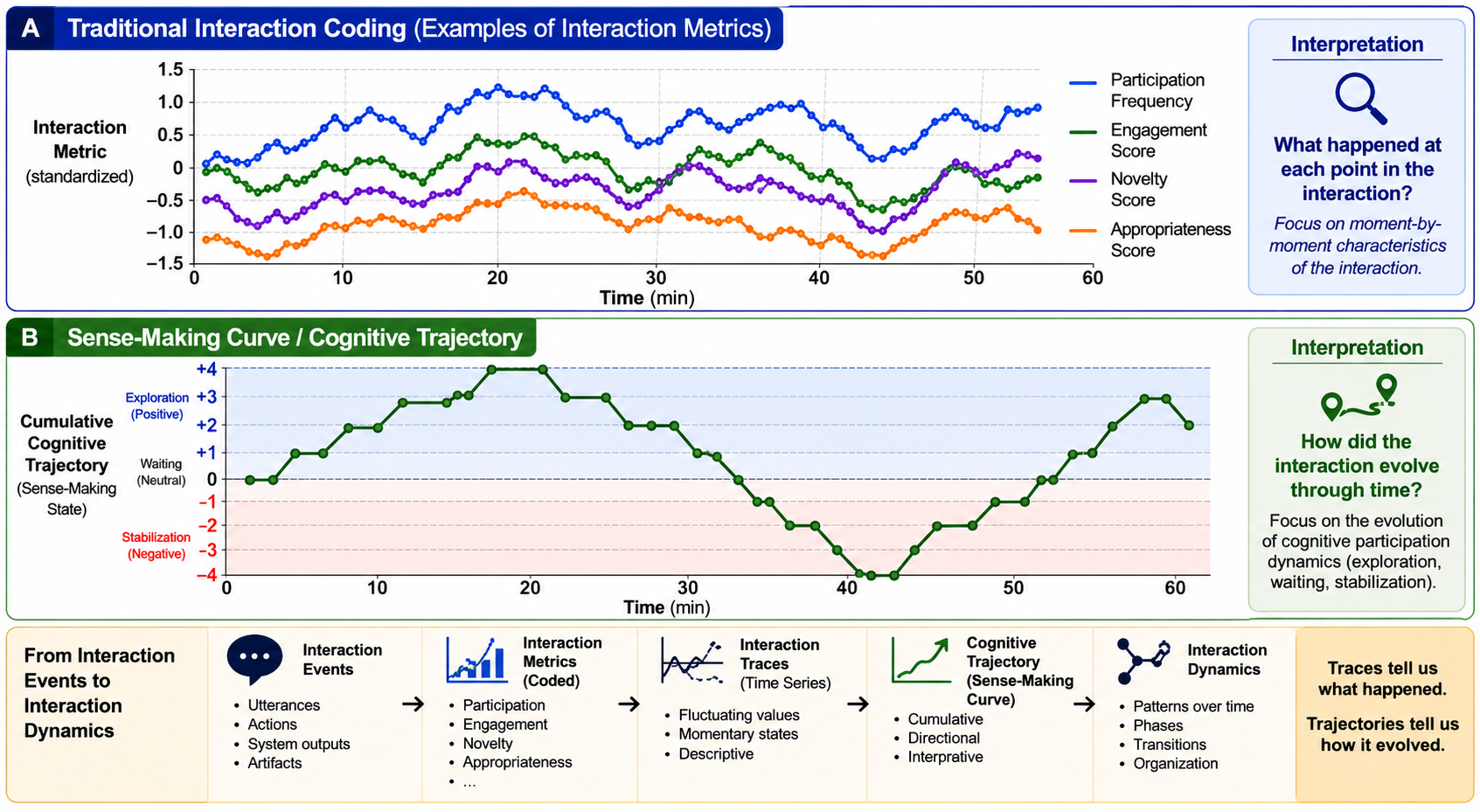}
  \caption{\textbf{Interaction traces versus cognitive trajectories.}
\normalfont{Interaction traces describe observable characteristics through time, whereas cognitive trajectories model how interaction dynamics and sense-making evolve. The figure illustrates the Cognitive Trajectory Principle: interpretation depends on the structure and meaning of the underlying state space, not merely the appearance of the resulting curve.}}

\Description{
Comparison between traditional interaction coding approaches and Cognitive Trajectory Modeling (CTM). The figure contrasts interaction traces, which describe what happened during an interaction, with cognitive trajectories, which describe how participation evolved through time.

Panel A, Traditional Interaction Coding, presents several example interaction metrics plotted as time series over a sixty-minute interaction. Metrics include participation frequency, engagement score, novelty score, and appropriateness score. Each metric fluctuates over time, producing descriptive traces that capture moment-by-moment properties of the interaction. These traces indicate what occurred at specific points in the interaction and support the analysis of observable characteristics, behaviors, and coded interaction measures. The accompanying interpretation emphasizes a focus on point-by-point interaction properties.

Panel B, Sense-Making Curve or Cognitive Trajectory, presents a cumulative trajectory derived from directional cognitive participation states. The vertical axis represents cognitive participation ranging from stabilization and clamped action at negative values, through waiting and neutral participation at zero, to exploratory engagement at positive values. The trajectory rises during periods of exploration, remains stable during sustained participation modes, and falls during periods of stabilization or commitment. Unlike the interaction traces above, the curve is interpreted as movement through a cognitive participation state space. The accompanying interpretation emphasizes understanding how interaction evolved through time rather than describing isolated events.

The lower section illustrates the analytical progression from interaction events to interaction dynamics. Interaction begins with events such as utterances, actions, system outputs, and artifacts. These events may be coded into interaction metrics including participation, engagement, novelty, and appropriateness. The resulting time-series traces describe fluctuating values and momentary states. CTM then transforms directional participation states into a cognitive trajectory or Sense-Making Curve, producing a cumulative, directional, and interpretable representation of participation through time. Analysis of the trajectory reveals higher-order interaction dynamics such as recurring patterns, phases, transitions, and organizational structure.

The central distinction emphasized by the figure is that interaction traces and cognitive trajectories are not equivalent. Traces describe what happened during an interaction, whereas trajectories describe how participation evolved through time. Cognitive trajectories therefore provide a framework for interpreting the temporal organization of cognition and interaction rather than merely describing moment-by-moment interaction characteristics.
}
\label{fig:traceVStrajectory}
\end{figure*}

\subsection{\textbf{Interaction Traces and Cognitive Trajectories}}

While many methods generate temporal representations of interaction, not all temporal representations support the same forms of interpretation. A useful distinction can therefore be drawn between interaction traces and cognitive trajectories. Interaction traces represent observable characteristics, events, behaviors, or metrics through time. They describe what occurred during an interaction and provide valuable information concerning participation, engagement, novelty, coordination, or other measurable properties. Cognitive trajectories represent evolving interaction dynamics through movement across directional cognitive attractor landscapes. Rather than describing isolated observations, trajectories model how interaction unfolds, stabilizes, reorganizes, and evolves through time. This distinction is important because temporal representations may appear visually similar while supporting fundamentally different theoretical interpretations. The Cognitive Trajectory Principle suggests that trajectory interpretation depends not upon visual appearance alone but upon the directional cognitive structure of the underlying attractor landscape.

\begin{table*}[t]
\centering
\caption{Interaction Traces versus Cognitive Trajectories}
\label{tab:traces_vs_trajectories}

\renewcommand{\arraystretch}{1.2}

\begin{tabularx}{\linewidth}{
>{\bfseries}p{2.6cm}
X
X
}
\toprule
Feature & Interaction Traces & Cognitive Trajectories \\
\midrule

Unit
&
Observable events
&
Directional cognitive states
\\

Focus
&
What happened
&
How interaction evolved
\\

Temporal Logic
&
Temporal description
&
Attractor landscape movement
\\

Interpretation
&
Interaction characteristics
&
Interaction dynamics
\\

Goal
&
Characterize activity
&
Model organization through time
\\

\bottomrule
\end{tabularx}
\end{table*}

\subsection{\textbf{The Dynamical Properties of Cognitive Trajectory Modeling}}

Although Cognitive Trajectory Modeling was initially introduced through directional participation states and Sense-Making Curves, the framework can also be understood through the language of dynamical systems theory. From this perspective, cognition and interaction do not simply move between discrete states. Rather, they organize around relatively stable regions of activity that persist across time. These regions may be conceptualized as attractors within an evolving organizational landscape.

Within Creative Sense-Making, clamped participation can be interpreted as a form of attractor stabilization. During clamped periods, cognition becomes committed to a particular course of action, interpretation, strategy, or organizational pattern. Attention narrows, behavior stabilizes, and the cognitive system becomes increasingly constrained by the current mode of engagement. In dynamical terms, the system has settled into an attractor basin.

Unclamped participation can be understood as a process of attractor destabilization and exploration. Rather than remaining committed to a single organizational pattern, the cognitive system becomes increasingly sensitive to alternative affordances, interpretations, and possibilities. This process often requires effort because stable attractors resist change. Shifting away from a well-established interpretation, habit, strategy, or interaction pattern may require the system to overcome the stabilizing forces that maintain the current attractor.

From this perspective, cognitive trajectories do not simply represent movement between predefined states. They represent movement through an evolving attractor landscape. Periods of stability correspond to attractor occupancy, whereas transitions correspond to processes of destabilization, exploration, reorganization, and re-stabilization. Trajectories therefore capture both persistence and change within cognitive and interactional systems.

This interpretation naturally extends CTM beyond the directional participation states of Creative Sense-Making. Future trajectory-oriented approaches may model the emergence, persistence, interaction, and transformation of attractors across multiple levels of organization. Cognitive attractors may give rise to interaction attractors such as recurring coordination patterns, collaborative routines, negotiation structures, or regulatory regimes. At broader scales, domain-level attractors may emerge in the form of stable creative strategies, design patterns, aesthetic conventions, or recurring solution pathways.

Viewed from this perspective, Cognitive Trajectory Modeling becomes a framework for studying how cognitive and interactional systems move through evolving attractor landscapes. Trajectories represent the paths taken through these landscapes, while attractors provide the organizational structures that shape, constrain, and guide the evolution of cognition, interaction, and creative activity through time.

\subsection{\textbf{Core Characteristics of Cognitive Trajectory Modeling}}

While future trajectory-based approaches may vary substantially in implementation, Cognitive Trajectory Modeling can be defined through several shared characteristics.

\subsubsection{\textbf{Directional State Representations}}

CTM frameworks employ coding states that possess directional meaning. The values do not merely represent magnitude, frequency, or observational assessment. Instead, they correspond to theoretically meaningful orientations within a cognitive attractor landscape. The state space must therefore possess internal structure. Movement between attractors reflects changes in modes of engagement rather than changes in quantity alone. This directional structure provides the foundation necessary for trajectory interpretation. Without directional states, cumulative movement cannot be interpreted as movement through a cognitive attractor landscape.

\subsubsection{\textbf{Cumulative Trajectory Generation}}

CTM frameworks transform directional participation states into cumulative trajectories across time. Importantly, accumulation is not used merely as a descriptive visualization technique. It serves as a mechanism for revealing larger patterns of cognitive organization that are difficult to observe at the level of isolated interaction events. The trajectory provides access to structures that emerge only through temporal unfolding. Exploration, stabilization, adaptation, oscillation, and regulation become visible because the interaction is modeled as an evolving pathway rather than a sequence of disconnected observations.

\subsubsection{\textbf{Temporal Organization}}

A central assumption of Cognitive Trajectory Modeling is that cognition possesses temporal structure. Participation unfolds through time. Consequently, understanding cognition requires methods capable of representing temporal organization rather than merely summarizing outcomes. CTM frameworks therefore focus on questions such as:

\begin{itemize}
    \item How do participation states evolve?
    \item How do interaction modes accumulate?
    \item How do exploratory and stabilizing dynamics reorganize?
    \item How do collaborative systems move between cognitive regimes?
\end{itemize}
These questions concern organization across time rather than isolated behavioral moments. The trajectory functions as a representation of this temporal structure.

\subsubsection{\textbf{Regulation Dynamics}}

CTM approaches also emphasize regulation. Creative interaction rarely involves static participation patterns. Participants continuously adjust their engagement in response to environmental conditions, collaborators, emerging constraints, and evolving opportunities for action. As a result, interaction often involves ongoing regulation between different participation modes.

Exploration may increase when uncertainty rises. Stabilization may emerge when promising directions are identified. Participants may reorganize their behavior when collaboration becomes ineffective. These regulatory dynamics often constitute a major component of creative activity. Trajectory-based approaches provide a means of representing such processes directly. Rather than treating regulation as a secondary phenomenon, CTM frameworks position regulation as a central organizational principle.

\subsubsection{\textbf{Oscillatory Structure}}

Many forms of cognition exhibit oscillatory organization. Participants move repeatedly between exploration and exploitation, divergence and convergence, environmental inspection and action commitment. Creative processes frequently emerge through these oscillatory dynamics. CTM frameworks are particularly suited to studying such phenomena because trajectories preserve the temporal relationships between participation states. Oscillations become visible as structural properties of the trajectory itself. Rather than reducing interaction to aggregate quantities, trajectory analysis allows researchers to examine how cycles of engagement emerge, stabilize, and reorganize across time. This makes CTM particularly useful for studying creative interaction, collaborative adaptation, and participatory sense-making.

\subsection{\textbf{Cognitive, Interaction, and Domain Dynamics}}

\begin{figure*}[!t]
  \centering
  \includegraphics[height=.9\textheight]{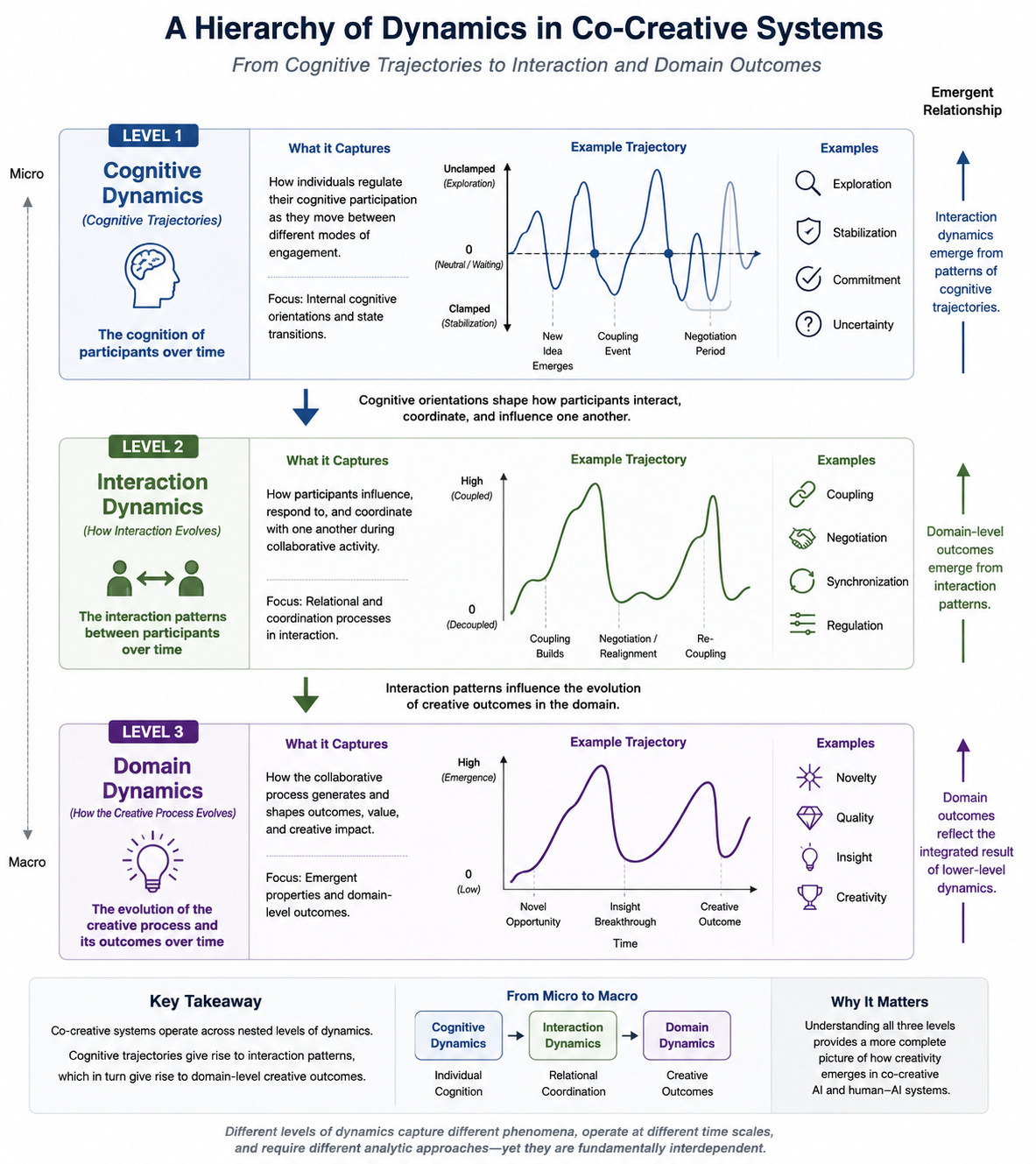}
  \caption{\textbf{A hierarchy of dynamics in co-creative systems developed by \cite{davis2025unlocking, davis2025ai}.}
\normalfont{Cognitive dynamics describe changing patterns of engagement, interaction dynamics describe coordination and regulation among participants, and domain dynamics describe the evolution of creative outcomes. The figure shows how cognitive trajectories contribute to interaction dynamics, which in turn shape domain-level outcomes across interdependent scales.}}

  \Description{
Conceptual diagram illustrating a three-level hierarchy of dynamics in co-creative systems, progressing from individual cognition to interaction processes and ultimately to creative outcomes. The hierarchy is organized from micro-level cognitive dynamics at the top to macro-level domain dynamics at the bottom, showing how higher-level phenomena emerge from lower-level processes through time.

Level 1, Cognitive Dynamics, represents cognitive trajectories within individual participants. This level captures how individuals regulate their cognitive participation as they move between different modes of engagement. An example trajectory oscillates between exploratory and stabilized states, passing through periods of idea emergence, coupling events, and negotiation. Example cognitive states include exploration, stabilization, commitment, and uncertainty. The figure emphasizes that cognitive trajectories describe how individuals participate in and regulate creative activity through time. A note on the right explains that interaction dynamics emerge from patterns of cognitive trajectories.

Level 2, Interaction Dynamics, represents how participants influence, coordinate with, and respond to one another during collaboration. An example trajectory depicts changing levels of coupling between participants, including periods of increasing coordination, negotiation or realignment, and re-coupling. Example interaction processes include coupling, negotiation, synchronization, and regulation. This level focuses on relational and coordination processes that emerge through interaction. The figure emphasizes that interaction patterns arise from cognitive trajectories and subsequently shape domain-level outcomes.

Level 3, Domain Dynamics, represents the evolution of the creative process and its outcomes. This level captures how collaborative processes generate value, novelty, insight, and creative products. An example trajectory shows the emergence of a novel opportunity, an insight breakthrough, and a creative outcome. Example domain-level phenomena include novelty, quality, insight, and creativity. This level focuses on emergent properties and outcomes that arise from collaborative activity. A note on the right explains that domain outcomes reflect the integrated result of lower-level cognitive and interaction dynamics.

Arrows between levels indicate an emergent relationship. Cognitive trajectories shape interaction patterns, interaction patterns influence the evolution of creative outcomes, and domain-level phenomena emerge from the combined effects of cognitive and interaction dynamics. The hierarchy therefore links individual cognition, collaborative interaction, and creative outcomes into a single explanatory framework.

A summary section at the bottom presents the central argument of the figure. Co-creative systems operate across nested levels of dynamics. Cognitive dynamics describe individual cognition, interaction dynamics describe relational coordination, and domain dynamics describe creative outcomes. Understanding all three levels provides a more complete account of how creativity emerges in human-AI and co-creative systems. The figure argues that different levels of dynamics capture different phenomena, operate at different temporal scales, and require different analytic approaches, yet remain fundamentally interdependent.
}
\label{fig:trajectories}
\end{figure*}

The distinction between participation dynamics and interaction dynamics helps clarify the scope of Cognitive Trajectory Modeling. Early work on Creative Sense-Making focused primarily on cognitive participation dynamics, examining how individuals move between exploratory, stabilizing, and action-oriented modes of engagement during creative interaction \cite{davis2024creative}. Within this perspective, trajectories represent changing patterns of cognitive participation through time. However, cognitive participation represents only one level within a broader hierarchy of dynamics. Interactions also exhibit emergent patterns of coupling, coordination, negotiation, synchronization, regulation, and adaptation that arise through relationships between participants, artifacts, and environments. These interaction dynamics cannot always be reduced to the cognitive trajectories of individual participants, although they emerge from and remain shaped by them. At an even broader level, interactions give rise to domain dynamics associated with the evolution of creative outcomes themselves. Processes such as novelty generation, insight formation, quality development, and creative emergence unfold across time as a consequence of underlying interaction patterns. From this perspective, Cognitive Trajectory Modeling can be understood as a framework for studying dynamics across multiple levels of organization. Cognitive trajectories provide one foundation for understanding interaction dynamics, which in turn shape the emergence of domain-level creative outcomes. Figure \ref{fig:trajectories} illustrates this hierarchical relationship.

\subsection{\textbf{Future Directions for Cognitive Trajectory Modeling}}

While Creative Sense-Making represents one early implementation of Cognitive Trajectory Modeling, the CTM framework is not limited to the directional participation states employed by CSM. More generally, CTM can be understood as a cognitive theory of interaction dynamics concerned with modeling trajectories through cognitively meaningful attractor landscapes. Future trajectory-oriented approaches may therefore employ substantially richer representations of cognitive organization.

One possibility involves multidimensional participation spaces. Rather than representing participation along a single exploratory stabilizing dimension, future models may track multiple dimensions simultaneously, including uncertainty, engagement, novelty seeking, coordination, commitment, or regulatory stability. Cognitive trajectories would then be represented as paths through higher-dimensional participation landscapes.

A second direction involves attractor-based representations derived from dynamical systems theory. Cognitive participation may not simply move between predefined states but instead organize around recurring regions of stability. Such attractors may correspond to persistent modes of engagement, collaborative routines, creative strategies, or stable interaction patterns. Trajectory analysis would then focus on how participants enter, remain within, and transition between attractor regions over time.

Probabilistic trajectory models provide another extension. Rather than assigning participants to discrete cognitive states, future frameworks may represent participation as probability distributions across multiple competing orientations. Such models may be particularly useful for studying ambiguous, transitional, or mixed participation states in which cognition exhibits partial commitment to multiple modes simultaneously. Trajectories would therefore describe evolving probability landscapes rather than movement between fixed categories.

Trajectory-based approaches may also incorporate regulatory mechanisms. The Enactive Model of Creativity conceptualized cognition as continuous regulation between exploratory and action-oriented modes of engagement. Future CTM systems could generalize this insight by modeling how cognitive systems regulate uncertainty, novelty, coherence, commitment, environmental coupling, or collaborative coordination through time. From this perspective, trajectories become representations of adaptive regulation rather than simple movement between states.

The concept of perceptual logic introduces another potential trajectory space. Early work on the Creative Trajectory Monitor proposed that creative systems may shift between local, regional, and global perceptual organizations depending upon the evolving interaction context. Future CTM frameworks may therefore model trajectories through perceptual logic spaces, tracking how attention and interpretation reorganize across multiple scales of analysis.

Taken together, these directions suggest that Cognitive Trajectory Modeling should not be understood as a single method but as a broader family of approaches concerned with representing cognition as movement through evolving attractor landscapes. The specific form of the landscape may vary. What remains consistent is the commitment to understanding cognition, creativity, and collaboration as temporally organized processes whose structure emerges through trajectories rather than isolated observations.

The significance of CTM therefore extends beyond Creative Sense-Making itself. The framework provides a general methodological perspective for studying interaction dynamics, adaptive participation, co-creative processes, and evolving forms of cognitive organization. Rather than asking how many behaviors occurred or how strongly a characteristic was expressed, CTM asks how cognitive participation moved, transformed, stabilized, and reorganized throughout the interaction process.

\subsection{\textbf{Cognitive Trajectory Modeling as a Research Program}}

The Cognitive Trajectory Principle reveals a broader methodological opportunity. Rather than viewing interaction coding solely as a mechanism for categorizing observable behaviors, researchers can treat interaction as a dynamic process whose organization emerges across time. This shift moves the focus from observation to trajectory. From description to organization. From isolated events to evolving participation dynamics.

Cognitive Trajectory Modeling represents an attempt to formalize this perspective. It identifies a family of approaches concerned with representing how cognition unfolds through interaction rather than merely documenting the observable products of interaction. From this perspective, the significance of Creative Sense-Making lies not only in the specific framework itself. Its larger contribution may be the introduction of a trajectory-based approach to modeling cognitive participation. It suggests a broader methodological direction for the future study of co-creative interaction, interaction-centered intelligence, and the temporal dynamics of participatory cognition. The trajectory becomes not merely a visualization of interaction. It becomes a representation of cognition unfolding through time.

\section{\textbf{Implications for Co-Creative AI}}

The distinction between interaction traces and cognitive trajectories extends beyond the interpretation of interaction curves. More broadly, it raises fundamental questions concerning how co-creative AI systems should be evaluated, what forms of interaction data should be considered meaningful, and what kinds of phenomena current evaluation frameworks are capable of capturing. Across computational creativity, human-computer interaction, creativity support tools, and co-creative AI research, evaluation methods increasingly rely upon observable interaction measures such as participation frequency, turn-taking counts, novelty ratings, appropriateness assessments, engagement metrics, behavioral annotations, and subjective evaluations of interaction quality \cite{davis2026human}.

These approaches have contributed valuable tools for studying collaborative systems. However, when co-creative interaction is analyzed primarily through observable ratings and behavioral measurements, important dimensions of the interaction process risk disappearing from analysis altogether. The distinction between cognitive trajectory modeling and observable interaction coding therefore has implications at a theoretical and methodological level. It raises broader questions concerning what co-creative AI research is actually measuring.

\subsection{\textbf{The Limits of Observable Evaluation}}

Many contemporary evaluation frameworks assume that meaningful information about co-creative interaction can be captured through observable characteristics of behavior.

Researchers may evaluate:

\begin{itemize}
    \item how frequently a participant contributes,
    \item how novel generated outputs appear,
    \item how appropriate suggestions are judged to be,
    \item how often turn exchanges occur,
    \item how engaged participants report feeling,
    \item how successfully collaborative goals are achieved.
\end{itemize}
These measures can provide useful information about interaction outcomes and observable activity. However, they do not necessarily reveal how interaction unfolds as a cognitive process. Two collaborations may exhibit identical participation counts while involving radically different forms of engagement. Likewise, two interactions may generate artifacts with comparable novelty ratings while emerging through entirely different participation dynamics. One interaction may involve sustained exploration, experimentation, and adaptive reorganization. Another may involve rigid execution, repetitive action, and minimal cognitive flexibility. Observable outcome measures may fail to distinguish between them. This limitation emerges because observable metrics often describe what occurred without modeling how participation evolved through time. As a result, the larger organizational dynamics of co-creative interaction remain difficult to observe.

\subsection{\textbf{What Disappears When Interaction Is Reduced to Observable Metrics}}

When interaction analysis focuses primarily on observable ratings, several important dimensions of co-creative activity become increasingly difficult to detect.

\subsubsection{\textbf{Exploratory Dynamics}}

Creative activity often involves periods of searching, probing, experimentation, uncertainty, and environmental inspection. These exploratory dynamics are frequently essential for the emergence of novel ideas. However, exploration may not always appear highly productive when evaluated through immediate outcomes. Participants may generate unsuccessful actions, incomplete structures, or abandoned pathways. Observable rating systems often struggle to distinguish exploratory engagement from inefficiency because both may appear similar at the behavioral level. As a result, important exploratory processes can disappear from analysis.

\subsubsection{\textbf{Stabilization Dynamics}}

Creative interaction also requires periods of commitment, execution, convergence, and stabilization. Ideas must eventually become actions. Possibilities must eventually become structures. Observable evaluation frameworks often register the existence of successful outputs but may fail to capture how stabilization emerges through interaction. The transition from exploration to commitment is itself a significant cognitive event. Yet it often remains invisible within static observational metrics.

\subsubsection{\textbf{Oscillatory Organization}}

Many creative processes do not proceed linearly. Participants frequently move back and forth between exploratory and stabilizing modes of engagement. Periods of divergence may be followed by convergence. Moments of commitment may be interrupted by renewed exploration. These oscillatory dynamics often play a central role in creative emergence. However, observable ratings rarely model these transitions directly. Instead, they tend to evaluate isolated moments or aggregate quantities. The organizational structure of the oscillation itself disappears.

\subsubsection{\textbf{Collaborative Regulation}}

Co-creative interaction frequently involves the regulation of participation between collaborators. Participants adjust their behavior in response to one another. They negotiate control, redistribute attention, coordinate exploration, stabilize emerging structures, and reorganize interaction patterns when collaboration becomes unproductive. These forms of regulation are often distributed across time and emerge through evolving participation dynamics. Observable metrics may capture isolated behaviors while missing the larger regulatory structures governing the interaction. The result is a loss of information concerning how collaboration actually sustains itself.

\subsection{\textbf{The Problem of Static Evaluation}}

A deeper issue emerges when observable interaction measures are treated as primary indicators of co-creative performance. Many such measures are fundamentally static. They describe characteristics of interaction moments. They do not necessarily model the temporal organization of interaction itself. This creates a methodological asymmetry. Co-creative systems are dynamic. Their evaluation frameworks are often comparatively static. As a result, evaluation procedures frequently focus on observable states while neglecting the trajectories through which those states emerge. The interaction becomes fragmented into isolated observations. The larger organizational process disappears. This issue becomes increasingly important as co-creative systems become more adaptive, participatory, and interactive. The more intelligence emerges through interaction, the less sufficient isolated observational metrics become.

\section{\textbf{Toward a Trajectory Science of Cognition and Interaction}}

The Cognitive Trajectory Principle provides a theoretical foundation for distinguishing cognitive trajectories from other forms of temporal representation. However, the broader significance of Cognitive Trajectory Modeling (CTM) extends beyond the interpretation of trajectories themselves. If cognition, interaction, and creative activity are organized through evolving trajectories rather than isolated events, then cognitive trajectories may provide a new foundation for understanding, designing, and evaluating co-creative systems.

Traditional approaches to human-computer interaction and co-creative AI frequently focus on observable events, behavioral responses, interaction metrics, or task outcomes. Systems typically adapt to immediate user actions, preferences, commands, or performance measures. While such approaches can support effective interaction, they often lack a representation of how participation evolves through time. Consequently, adaptation frequently occurs in response to local interaction events rather than the broader cognitive and interactional context in which those events occur.

From the perspective of Cognitive Trajectory Modeling, the trajectory itself becomes an important source of information. A cognitive trajectory provides a temporally organized representation of how a participant is exploring possibilities, stabilizing ideas, transitioning between modes of engagement, responding to uncertainty, reorganizing strategies, and regulating participation across time. This information may allow co-creative systems to move beyond event-based interaction toward forms of adaptation grounded in the evolving organization of participation itself.

The following hypotheses illustrate several directions through which Cognitive Trajectory Modeling may contribute to the development of trajectory-aware co-creative systems and interaction-centered approaches to artificial intelligence.

\subsection{\textbf{H1: Event-Based Adaptation Is Insufficient for Sustained Co-Creation}}

Traditional interactive systems frequently respond to the most recent observable action performed by a user. In contrast, trajectory-aware systems maintain a representation of how interaction has evolved across time and use that trajectory to interpret current behavior. CTM predicts that systems capable of modeling cognitive trajectories will outperform systems that rely solely upon event-level interaction data. The reason is that individual actions are often ambiguous when considered in isolation. The same behavior may possess different meanings depending upon the broader trajectory within which it occurs. A suggestion, drawing stroke, pause, revision, or conversational contribution may reflect exploration, stabilization, uncertainty, reorganization, or commitment depending upon its trajectory context. This hypothesis predicts that trajectory-aware systems will exhibit improved collaboration quality, user engagement, perceived intelligence, and creative support compared to systems that respond only to local interaction events.

\subsection{\textbf{H2: Cognitive Trajectories Provide Context for Adaptive AI Behavior}}

Most adaptive systems modify their behavior using measures such as user preferences, performance metrics, engagement scores, or task outcomes. While useful, these measures often provide limited information concerning the participant’s current cognitive organization. CTM predicts that adaptation should be informed by trajectory position rather than observable behavior alone. Different regions of a trajectory correspond to different cognitive contexts and therefore may require different forms of support. For example, highly exploratory trajectories may benefit from increased novelty, alternative perspectives, and divergent suggestions. In contrast, stabilizing trajectories may benefit from refinement, elaboration, evaluation, and support for commitment. Transitional trajectories may benefit from interventions that help facilitate reorganization or clarify emerging directions. Trajectory context improves adaptive decision making. This hypothesis predicts that adaptive systems capable of responding to trajectory context will provide more effective support than systems that adapt solely to immediate interaction events or static user models.

\subsection{\textbf{H3: Successful Co-Creation Depends Upon Trajectory Coordination}}

Traditional evaluations of co-creative systems often focus on properties of outputs, including novelty, quality, appropriateness, or creativity. While these measures remain important, CTM suggests that successful collaboration may depend equally upon how effectively a system coordinates with the evolving trajectory of the human participant. From this perspective, co-creation is not simply the production of artifacts but the regulation of a shared interaction process. Effective collaboration may therefore depend upon whether the AI appropriately complements, supports, challenges, or extends the participant’s current mode of engagement. This hypothesis predicts that successful collaborations will exhibit higher levels of trajectory coordination than unsuccessful collaborations. Such coordination may involve synchronized transitions, complementary participation patterns, adaptive co-regulation, or mutual responsiveness to changing interactional conditions. Conversely, collaboration breakdowns may emerge when the system persistently acts against the participant’s evolving trajectory.

\subsection{\textbf{H4: Trajectory-Based Adaptation Produces More Human-Like Collaboration}}

Many interactive systems operate through reactive stimulus-response relationships. User actions trigger system responses, which in turn trigger additional user actions. Although effective for many tasks, such interactions often lack the temporal continuity associated with human collaboration. CTM predicts that systems capable of modeling cognitive trajectories will produce interactions that appear more collaborative, anticipatory, and contextually aware. Rather than responding solely to what has just occurred, trajectory-aware systems respond to where the interaction appears to be going. This hypothesis predicts that users interacting with trajectory-aware systems will report greater perceptions of understanding, partnership, collaboration quality, and social presence than users interacting with event-based systems. More broadly, trajectory-aware interaction may represent an important step toward interaction-centered forms of artificial intelligence that participate in evolving processes rather than merely responding to isolated inputs. Users may perceive trajectory-aware adaptation as more collaborative and intelligent.

\subsection{\textbf{H5: Drift Signals Opportunities for Creative Intervention}}

A central implication of trajectory-based analysis is that deviations from established participation patterns may contain important information about the evolving state of an interaction. Whereas traditional systems may treat such deviations as noise, CTM suggests that drift may serve as an indicator of emerging reorganization. Periods of increasing drift may reflect growing uncertainty, instability within existing attractors, dissatisfaction with current directions, emerging opportunities, or the early stages of creative transition. Rather than suppressing these deviations, trajectory-aware systems may benefit from monitoring and responding to them. This hypothesis predicts that interventions informed by measures of trajectory drift will produce more effective creative support than interventions triggered at fixed intervals or through purely reactive strategies. Such interventions may include introducing novelty, proposing alternative directions, encouraging exploration, facilitating reflection, or supporting transitions between participation regimes.

Collectively, these hypotheses reposition Cognitive Trajectory Modeling from a framework for analyzing interaction dynamics to a foundation for designing future co-creative systems. Rather than treating cognition and interaction as collections of isolated events, CTM proposes that participation unfolds through trajectories moving across cognitively meaningful attractor landscapes. If this perspective is correct, then future intelligent systems may benefit not only from understanding what users are doing, but also from understanding where their cognitive trajectories have been, where they currently reside, and where they may be heading next. Such systems would represent a shift from event-centered interaction toward trajectory-centered participation, providing a foundation for interaction-centered intelligence, adaptive co-creation, and future generations of human-AI collaboration.

\section{\textbf{Discussion}}

This paper has argued that the interpretation of temporal interaction representations depends upon the structure of the underlying state space from which those representations are generated. The Cognitive Trajectory Principle proposed here suggests that temporal representations are only interpretable as cognitive trajectories when their underlying states possess directional cognitive meaning. This distinction shifts attention away from the visual appearance of temporal curves and toward the theoretical assumptions embedded within the coding systems that generate them.

From this perspective, the central contribution of Cognitive Trajectory Modeling (CTM) is not the introduction of a particular coding scheme, but the identification of a broader methodological category concerned with modeling evolving participation dynamics through time. CTM provides a framework for distinguishing between temporal representations that describe observable characteristics and those that model movement through theoretically meaningful cognitive attractor landscapes. The distinction is important because different forms of temporal representation support different forms of interpretation, even when their visual structures appear similar.

Throughout this paper, we have argued that the interpretability of cumulative trajectories depends not on accumulation alone but on the structure of the underlying state space. The Cognitive Trajectory Principle proposed here states that: \textbf{A cumulative trajectory is only theoretically interpretable as a cognitive trajectory when the underlying coding states possess directional cognitive meaning.} This principle provides a way of distinguishing between cumulative representations that model evolving participation dynamics and those that merely aggregate observational measurements. Observable interaction measures can provide important information concerning participation, engagement, novelty, appropriateness, embodiment, coordination, and collaborative activity. The argument advanced here is that observational coding and cognitive trajectory modeling are different forms of analysis that should not be treated as theoretically interchangeable. The distinction becomes increasingly important as co-creative AI research continues to expand.

This distinction motivated the introduction of Cognitive Trajectory Modeling (CTM) as a broader theoretical and methodological framework for understanding interaction dynamics. CTM encompasses a family of approaches concerned with how cognition, interaction, and creative processes unfold through time as trajectories across cognitively meaningful attractor landscapes. CTM provides a theoretical framework and methodological family for understanding how interaction dynamics emerge, stabilize, and evolve through time.

From this perspective, Creative Sense-Making can be understood as an early example of a larger trajectory-oriented research program rather than as an isolated framework. The significance of Creative Sense-Making may therefore extend beyond its original implementation. Its broader contribution lies in demonstrating how interaction can be modeled as an evolving trajectory of participation rather than as a collection of isolated observations. This shift has implications not only for co-creative AI, but also for interaction-centered intelligence, participatory cognition, distributed creativity, enactive systems, and future approaches to quantified interaction analysis.

Future trajectory-oriented approaches need not be restricted to the directional state spaces used within Creative Sense-Making. Multidimensional state spaces, attractor landscapes, probabilistic trajectories, perceptual logic trajectories, and other forms of temporal organization may all be incorporated within the broader CTM framework. The central requirement is not a particular coding scheme, but a theoretically meaningful state space capable of supporting trajectory interpretation. By focusing attention on the relationship between state-space structure and temporal representation, Cognitive Trajectory Modeling provides a foundation for future research on co-creative AI, interaction-centered intelligence, participatory cognition, distributed creativity, and adaptive collaborative systems. The challenge moving forward is not simply to measure interaction, but to develop methods capable of representing how interaction becomes organized through time.

\section{Conclusion}

As co-creative AI research increasingly shifts its attention from isolated outputs toward the dynamics of interaction itself, the need for methods capable of representing how participation evolves through time becomes increasingly important. This paper introduced Cognitive Trajectory Modeling (CTM) as a framework for understanding interaction as movement through cognitively meaningful attractor landscapes and proposed the Cognitive Trajectory Principle as a criterion for interpreting temporal representations of interaction. The Cognitive Trajectory Principle suggests that the interpretability of a temporal representation depends not upon its visual appearance alone, but upon the structure of the state space from which it is generated. Temporal representations become theoretically interpretable as cognitive trajectories when their underlying states possess directional cognitive meaning and support movement through a meaningful participation landscape. This shifts attention away from the accumulation of observations and toward the organization of participation through time. More broadly, this paper argues that creativity, collaboration, and intelligence are not merely collections of isolated interaction events. They are temporally organized processes that unfold through patterns of exploration, stabilization, adaptation, regulation, and reorganization across time. Understanding such processes requires methods capable of representing how participation changes, persists, and transforms within evolving interaction landscapes. As increasingly adaptive and collaborative AI systems emerge, the challenge facing co-creative AI research may no longer be simply measuring interaction, but understanding how interaction becomes organized through time. Cognitive Trajectory Modeling offers one step toward this goal by providing a framework for studying participation as a dynamic process of evolving trajectories rather than a collection of isolated observations. In doing so, it opens the possibility of a broader trajectory-based science of cognition, creativity, and collaborative intelligence.

\section{Acknowledgments}

The author would like to acknowledge the collaborative role of ChatGPT (OpenAI) in the development of this manuscript. Through extensive iterative dialogue, critical discussion, conceptual refinement, editorial feedback, literature synthesis, and theoretical exploration, ChatGPT served as a collaborative intellectual partner throughout the writing process. The core concepts, theoretical frameworks, and research program presented in this paper—including Creative Sense-Making, the Enactive Model of Creativity, the Cognitive Trajectory Principle, and Cognitive Trajectory Modeling—originate from the author's prior research. However, many aspects of the manuscript's organization, articulation, argumentation, and refinement emerged through ongoing human–AI collaboration during the development of the work.

\bibliographystyle{ACM-Reference-Format}
\bibliography{pubs}

@incollection{Davis2015EnactiveModel,
  author    = {Davis, Nicholas and Hsiao, Chih-Pin and Popova, Yulia and Magerko, Brian},
  title     = {An Enactive Model of Creativity for Computational Collaboration and Co-Creation},
  booktitle = {Creativity in the Digital Age},
  editor    = {Zagalo, Nelson and Branco, Pedro},
  publisher = {Springer},
  year      = {2015},
  pages     = {109--133},
  doi       = {10.1007/978-1-4471-6681-8_7}
}

@inproceedings{Fischer2005SocialCreativity,
  author    = {Fischer, Gerhard},
  title     = {Distances and Diversity: Sources for Social Creativity},
  booktitle = {Proceedings of the Fifth Conference on Creativity and Cognition},
  year      = {2005},
  pages     = {128--136},
  publisher = {ACM},
  doi       = {10.1145/1056224.1056243}
}

@inproceedings{Frich2019Landscape,
  author={Frich, Jonas and MacDonald Vermeulen, Lindsay and Remy, Christian and Biskjaer, Michael Mose and Dalsgaard, Peter},
  title     = {Mapping the Landscape of Creativity Support Tools in HCI},
  booktitle = {Proceedings of the 2019 CHI Conference on Human Factors in Computing Systems},
  year      = {2019},
  articleno = {389},
  pages     = {1--18},
  publisher = {ACM},
  doi       = {10.1145/3290605.3300619}
}

@book{clark1997being,
  author    = {Clark, Andy},
  title     = {Being There: Putting Brain, Body, and World Together Again},
  publisher = {MIT Press},
  address   = {Cambridge, MA},
  year      = {1997}
}

@inproceedings{magerko2009empirical,
  title={An empirical study of cognition and theatrical improvisation},
  author={Magerko, Brian and Manzoul, Waleed and Riedl, Mark and Baumer, Allan and Fuller, Daniel and Luther, Kurt and Pearce, Celia},
  booktitle={Proceedings of the seventh ACM conference on Creativity and cognition},
  pages={117--126},
  year={2009}
}

@incollection{davis2026human,
  title={Human-AI co-creation: A new interaction paradigm for human-AI interaction},
  author={Davis, Nicholas and Clemens, Michael and Rezwana, Jeba and Browne, Eric},
  booktitle={Handbook of Human-Centered Artificial Intelligence},
  pages={833--889},
  year={2026},
  publisher={Springer}
}

@incollection{davis2025unlocking,
  title={Unlocking the Black Box of artificial media with quantified and explainable co-creative AI systems},
  author={Davis, Nicholas and Clemens, Michael and Browne, Eric and Rezwana, Jeba},
  booktitle={Artificial Media: Emerging Trends in Narratives, Education and Creative Practice},
  pages={21--48},
  year={2025},
  publisher={Springer}
}

@book{finke1992creative,
  author    = {Finke, Ronald A. and Ward, Thomas B. and Smith, Steven M.},
  title     = {Creative Cognition: Theory, Research, and Applications},
  publisher = {MIT Press},
  address   = {Cambridge, MA},
  year      = {1992}
}

@collection{stewart2010enaction,
  editor    = {Stewart, John and Gapenne, Olivier and Di Paolo, Ezequiel A.},
  title     = {Enaction: Toward a New Paradigm for Cognitive Science},
  publisher = {MIT Press},
  address   = {Cambridge, MA},
  year      = {2010}
}

@article{davis2017quantifying,
  author    = {Davis, Nicholas and Hsiao, Chih-Pin and Singh, Kunwar Yashraj and Lin, Brenda and Magerko, Brian},
  title     = {Quantifying Collaboration with a Co-Creative Drawing Agent},
  journal   = {ACM Transactions on Interactive Intelligent Systems},
  volume    = {7},
  number    = {4},
  articleno = {25},
  pages     = {1--25},
  year      = {2017},
  publisher = {ACM},
  doi       = {10.1145/3002426}
}

@article{Shneiderman2007CreativitySupport,
  author  = {Shneiderman, Ben},
  title   = {Creativity Support Tools: Accelerating Discovery and Innovation},
  journal = {Communications of the ACM},
  volume  = {50},
  number  = {12},
  pages   = {20--32},
  year    = {2007},
  doi     = {10.1145/1323688.1323689}
}

@misc{Karimi2018EvaluatingCreativity,
  author        = {Karimi, Pegah and Grace, Kazjon and Maher, Mary Lou and Davis, Nicholas},
  title         = {Evaluating Creativity in Computational Co-Creative Systems},
  year          = {2018},
  eprint        = {1807.09886},
  archivePrefix = {arXiv},
  primaryClass  = {cs.AI}
}

@inproceedings{Davis2017CSM,
  author    = {Davis, Nicholas and Hsiao, Chih-Pin and Singh, Kunwar Yashraj and Lin, Brenda and Magerko, Brian},
  title     = {Creative Sense-Making: Quantifying Interaction Dynamics in Co-Creation},
  booktitle = {Proceedings of the 2017 ACM SIGCHI Conference on Creativity and Cognition},
  year      = {2017},
  pages      = {356--366},
  publisher = {ACM},
  doi       = {10.1145/3059454.3059478}
}

@inproceedings{winston2017turn,
  title={Turn-taking with improvisational co-creative agents},
  author={Winston, Lauren and Magerko, Brian},
  booktitle={Proceedings of the AAAI conference on artificial intelligence and interactive digital entertainment},
  volume={13},
  number={1},
  pages={129--135},
  year={2017}
}

@inproceedings{davis2015enactive,
  title={An enactive characterization of pretend play},
  author={Davis, Nicholas and Comerford, Margeaux and Jacob, Mikhail and Hsiao, Chih-Pin and Magerko, Brian},
  booktitle={Proceedings of the 2015 ACM SIGCHI Conference on Creativity and Cognition},
  pages={275--284},
  year={2015}
}

@inproceedings{davis2013human,
  title={Human-computer co-creativity: Blending human and computational creativity},
  author={Davis, Nicholas},
  booktitle={Proceedings of the AAAI conference on artificial intelligence and interactive digital entertainment},
  volume={9},
  number={6},
  pages={9--12},
  year={2013}
}

@inproceedings{davis2011computing,
  title={Computing harmony with PerLogicArt: perceptual logic inspired collaborative art},
  author={Davis, Nicholas and Do, Ellen Yi-Luen and Gupta, Pramod and Gupta, Shruti},
  booktitle={Proceedings of the 8th ACM conference on Creativity and cognition},
  pages={185--194},
  year={2011}
}

@article{davis2026interaction,
  title={Interaction-Centered Intelligence: Toward Interaction as the Primary Unit of Analysis in Co-Creative AI and Human-AI Systems},
  author={Davis, Nicholas},
  journal={arXiv preprint arXiv:2606.00807},
  year={2026}
}

@article{dorst2011core,
  title={The core of ‘design thinking’and its application},
  author={Dorst, Kees},
  journal={Design studies},
  volume={32},
  number={6},
  pages={521--532},
  year={2011},
  publisher={Elsevier}
}

@inproceedings{beaudouin2004designing,
  title={Designing interaction, not interfaces},
  author={Beaudouin-Lafon, Michel},
  booktitle={Proceedings of the working conference on Advanced visual interfaces},
  pages={15--22},
  year={2004}
}

@article{jordanous2016four,
  title={Four PPPPerspectives on computational creativity in theory and in practice},
  author={Jordanous, Anna},
  journal={Connection Science},
  volume={28},
  number={2},
  pages={194--216},
  year={2016},
  publisher={Taylor \& Francis}
}

@article{beer2000dynamical,
  title={Dynamical approaches to cognitive science},
  author={Beer, Randall D},
  journal={Trends in cognitive sciences},
  volume={4},
  number={3},
  pages={91--99},
  year={2000},
  publisher={Elsevier}
}

@article{davis2025ai,
  title={AI drawing partner: co-creative drawing agent and research platform to model co-creation},
  author={Davis, Nicholas and Rafner, Janet},
  journal={arXiv preprint arXiv:2501.06607},
  year={2025}
}

@inproceedings{bown2020speculative,
  title={A Speculative Exploration of the Role of Dialogue in Human-Computer Co-creation.},
  author={Bown, Oliver and Grace, Kazjon and Bray, Liam and Ventura, Dan},
  booktitle={ICCC},
  pages={25--32},
  year={2020}
}

@article{jo2026logs,
  title={From Logs to Agents: Reconstructing High-Level Creative Workflows from Low-Level Raw System Traces},
  author={Jo, Tae Hee and Hyun, Kyung Hoon},
  journal={arXiv preprint arXiv:2603.07609},
  year={2026}
}

@article{lin2023ontology,
  title={An ontology of co-creative AI systems},
  author={Lin, Zhiyu and Riedl, Mark},
  journal={arXiv preprint arXiv:2310.07472},
  year={2023}
}

@article{guzdial2019interaction,
  title={An interaction framework for studying co-creative ai},
  author={Guzdial, Matthew and Riedl, Mark},
  journal={arXiv preprint arXiv:1903.09709},
  year={2019}
}

@article{peng2026state,
  title={From State Changes to Creative Decisions: Documenting and Interpreting Traces Across Creative Domains},
  author={Peng, Xiaohan and Piliouras, Sotiris and Nujaim, Carl Abou Saada},
  journal={arXiv preprint arXiv:2603.07184},
  year={2026}
}

@inproceedings{Deshpande2023OCSM,
  author    = {Deshpande, Manoj and Trajkova, Milka and Knowlton, Andrea and Magerko, Brian},
  title     = {Observable Creative Sense-Making (OCSM): A Method for Quantifying Improvisational Co-Creative Interaction},
  booktitle = {Proceedings of the 2023 ACM Creativity and Cognition Conference},
  year      = {2023},
  publisher = {ACM},
    pages={103--115},
  doi       = {10.1145/3591196.3593514}
}

@incollection{DiPaolo2010Horizons,
  author    = {Di Paolo, Ezequiel A. and Rohde, Marieke and De Jaegher, Hanne},
  title     = {Horizons for the Enactive Mind: Values, Social Interaction, and Play},
  booktitle = {Enaction: Toward a New Paradigm for Cognitive Science},
  editor    = {Stewart, John and Gapenne, Olivier and Di Paolo, Ezequiel A.},
  publisher = {MIT Press},
  year      = {2010},
  pages     = {33--87}
}

@book{Hutchins1995,
  author    = {Hutchins, Edwin},
  title     = {Cognition in the Wild},
  publisher = {MIT Press},
  year      = {1995}
}

@inproceedings{Kantosalo2016Modes,
  author    = {Kantosalo, Anna and Toivonen, Hannu},
  title     = {Modes for Creative Human-Computer Collaboration},
  booktitle = {Proceedings of the Seventh International Conference on Computational Creativity},
    pages={77--84},
  year      = {2016}
}

@book{Sawyer2007GroupGenius,
  author    = {Sawyer, R. Keith},
  title     = {Group Genius: The Creative Power of Collaboration},
  publisher = {Basic Books},
  year      = {2007}
}

@book{thompson2010mind,
  title={Mind in life: Biology, phenomenology, and the sciences of mind},
  author={Thompson, Evan},
  year={2010},
  publisher={Harvard University Press}
}

@book{Varela1991EmbodiedMind,
  author    = {Varela, Francisco J. and Thompson, Evan and Rosch, Eleanor},
  title     = {The Embodied Mind: Cognitive Science and Human Experience},
  publisher = {MIT Press},
  year      = {1991}
}

@article{DeJaegher2007PSM,
author  = {De Jaegher, Hanne and Di Paolo, Ezequiel A.},
title   = {Participatory Sense-Making: An Enactive Approach to Social Cognition},
journal = {Phenomenology and the Cognitive Sciences},
year    = {2007},
volume  = {6},
number  = {4},
pages   = {485--507},
year={2007},
publisher={Springer}

}

@article{NewellSimon1976,
author    = {Allen Newell and Herbert A. Simon},
title     = {Computer Science as Empirical Inquiry: Symbols and Search},
journal   = {Communications of the ACM},
volume     = {19},
number     = {3},
pages      = {113--126},
year       = {1976},
doi        = {10.1145/360018.360022}
}

@article{Newell1980,
author    = {Allen Newell},
title     = {Physical Symbol Systems},
journal   = {Cognitive Science},
volume     = {4},
number     = {2},
pages      = {135--183},
year       = {1980},
doi        = {10.1207/s15516709cog0402_2}
}

@book{Simon1979,
author    = {Herbert A. Simon},
title     = {Models of Thought},
publisher = {Yale University Press},
address   = {New Haven, CT},
year      = {1979}
}

@article{Gabora2017,
  author  = {Liane Gabora},
  title   = {Honing Theory: A Complex Systems Framework for Creativity},
  journal = {Nonlinear Dynamics, Psychology, and Life Sciences},
  volume  = {21},
  number  = {1},
  pages   = {35--88},
  year    = {2017}
}

@book{Kelso1995,
  title={Dynamic patterns: The self-organization of brain and behavior},
  author={Kelso, JA Scott},
  year={1995},
  publisher={MIT press}
}

@book{ThelenSmith1994,
  author    = {Esther Thelen and Linda B. Smith},
  title     = {A Dynamic Systems Approach to the Development of Cognition and Action},
  publisher = {MIT Press},
  address   = {Cambridge, MA},
  year      = {1994}
}

@article{vanGelder1998,
  author  = {Tim van Gelder},
  title   = {The Dynamical Hypothesis in Cognitive Science},
  journal = {Behavioral and Brain Sciences},
  volume  = {21},
  number  = {5},
  pages   = {615--628},
  year    = {1998},
  doi     = {10.1017/S0140525X98001733}
}

@book{gibson2014ecological,
  title={The ecological approach to visual perception: classic edition},
  author={Gibson, James J},
  year={2014},
  publisher={Psychology press}
}

@book{noe2004action,
  title={Action in perception},
  author={No{\"e}, Alva},
  year={2004},
  publisher={MIT press}
}

@article{LairdNewellRosenbloom1987,
author    = {John E. Laird and Allen Newell and Paul S. Rosenbloom},
title     = {Soar: An Architecture for General Intelligence},
journal   = {Artificial Intelligence},
volume     = {33},
number     = {1},
pages      = {1--64},
year       = {1987},
doi        = {10.1016/0004-3702(87)90050-6}
}

@incollection{davis2024creative,
  title={Creative Sense-Making: A Cognitive Framework for Modelling Interaction Dynamics in Co-Creative AI},
  author={Davis, Nicholas},
  booktitle={Artificial Intelligence, Co-Creation and Creativity},
  pages={45--60},
  year={2024},
  publisher={Routledge}
}

@inproceedings{Fails2003,
author = {Fails, Jerry and Olsen, Dan},
title = {Interactive Machine Learning},
booktitle = {Proceedings of IUI},
year = {2003},
pages = {39--45}
}

@article{amershi2014,
  title={Power to the people: The role of humans in interactive machine learning},
  author={Amershi, Saleema and Cakmak, Maya and Knox, William Bradley and Kulesza, Todd},
  journal={AI magazine},
  volume={35},
  number={4},
  pages={105--120},
  year={2014}
}

@inproceedings{davis2014building,
  title={Building Artistic Computer Colleagues with an Enactive Model of Creativity.},
  author={Davis, Nicholas M and Popova, Yanna and Sysoev, Ivan and Hsiao, Chih-Pin and Zhang, Dingtian and Magerko, Brian},
  booktitle={ICCC},
  pages={38--45},
  year={2014}
}


\end{document}